\documentclass[%
  reprint,
groupedaddress,
 amsmath,amssymb,
 aps,
]{revtex4-1}

\usepackage{graphicx}

\usepackage{amsmath}
\usepackage{dcolumn}
\usepackage{bm}
\usepackage{comment}
\usepackage{braket}
\usepackage[mathlines]{lineno}

\newcommand{\Kcbulk}{K_{\mathrm{c}}^{\mathrm{bulk}}}
\newcommand{\nn}{\nonumber \\}
\newcommand{\tTr}{\mathrm{tTr}}
\newcommand{\cardy}[1]{\Ket{\tilde{{#1}}}}
\newcommand{\ydrac}[1]{\Bra{\tilde{{#1}}}}
\newcommand{\cardyw}[1]{\Ket{\widetilde{{#1}}}}

\newcommand{\eq}[1]{Eq.~(\ref{#1})}

\begin{document}


\title{Boundary conformal spectrum and surface critical behaviors\\of the classical spin systems: a tensor network renormalization study}


\author{Shumpei Iino}
\email[E-mail address: ]{iino@issp.u-tokyo.ac.jp}
\affiliation{Institute for Solid State Physics, The University of Tokyo, Kashiwa, Chiba, Japan}
\author{Satoshi Morita}
\affiliation{Institute for Solid State Physics, The University of Tokyo, Kashiwa, Chiba, Japan}
\author{Naoki Kawashima}
\affiliation{Institute for Solid State Physics, The University of Tokyo, Kashiwa, Chiba, Japan}


\date{\today}

\begin{abstract}
  We numerically obtain the conformal spectrum of several classical spin models on a two-dimensional lattice with open boundaries, for every boundary fixed point obtained by the Cardy's derivation~[J. L. Cardy, Nucl. Phys. B \textbf{324}, 581 (1989)]. In order to extract accurate conformal data, we implement the tensor network renormalization algorithm~[G. Evenbly and G. Vidal, Phys. Rev. Lett. \textbf{115}, 180405 (2015)] extended so as to be applicable to a square lattice with open boundaries. We successfully compute the boundary conformal spectrum consistent with the underlying boundary conformal field theories (BCFTs) for the Ising, tri-critical Ising, and 3-state Potts models on the lattice, which allows us to confirm the validity of the BCFT analyses for the surface critical behaviors of those lattice models.
\end{abstract}


\maketitle



\section{Introduction\label{sec:intro}}

Assumption of the conformal invariance is a powerful method for investigating critical phenomena, especially in two dimension whose universal structures have been revealed in detail~\cite{Belavin1984,Francesco_CFT}. Because the consistency with the emergent conformal invariance has been confirmed for many lattice systems by various methods such as exact results and numerical simulations, this enhanced symmetry in most of two-dimensional critical systems has been established beyond reasonable doubt.

Right after Belavin, Polyakov, and Zamolodchikov had established the basic theory of the two-dimensional conformal filed theory (CFT)~\cite{Belavin1984}, Cardy constructed CFT for the system with boundaries such as a semi-infinite plane~\cite{Cardy1984}, which is called boundary CFT (BCFT). In order to keep the boundaries invariant under the conformal transformations, the conformal symmetry is partially restricted in BCFT, which results in the difference from the ordinary CFT, such as an absence of antiholomorphic part of the Virasoro algebra. By considering the conditions for the conformal invariance of the boundaries, one can obtain the conformally invariant boundary conditions (b.c.'s) called Cardy states characterized by the primary fields, and also boundary fixed points~\cite{CARDY1989581}. In BCFT, the operator contents change depending on what b.c.'s are imposed on the boundaries~\cite{CARDY1986200}, which can be calculated by the fusion rules between the primary fields corresponding to the Cardy states.

When a system undergoes a phase transition, its boundaries also exhibit a singularity called surface critical behavior~\cite{Binder1983_DG}. In general, the thermodynamic free energy $F$ can be series-expanded in terms of the system size $L$ as
\begin{equation}
  \frac{F}{L^d} = f_{\mathrm{bulk}} + \frac{1}{L}f_{\mathrm{surface}} + \cdots,
\end{equation}
where $d$ is the spatial dimension. The singular part of $f_{\mathrm{surface}}$ leads to the singularity of physical quantities at the boundary, such as the surface magnetization, surface energy, and those derivatives. The critical exponents of the surface quantities are in general different from those of bulk ones; for instance the surface magnetic exponent of the two-dimensional Ising model is $\beta_1 = \frac{1}{2}$~\cite{McCoy1967}, while $\beta=\frac{1}{8}$ for the bulk magnetization as is well known~\cite{Yang1952}.

\begin{figure}
\includegraphics[width=7cm]{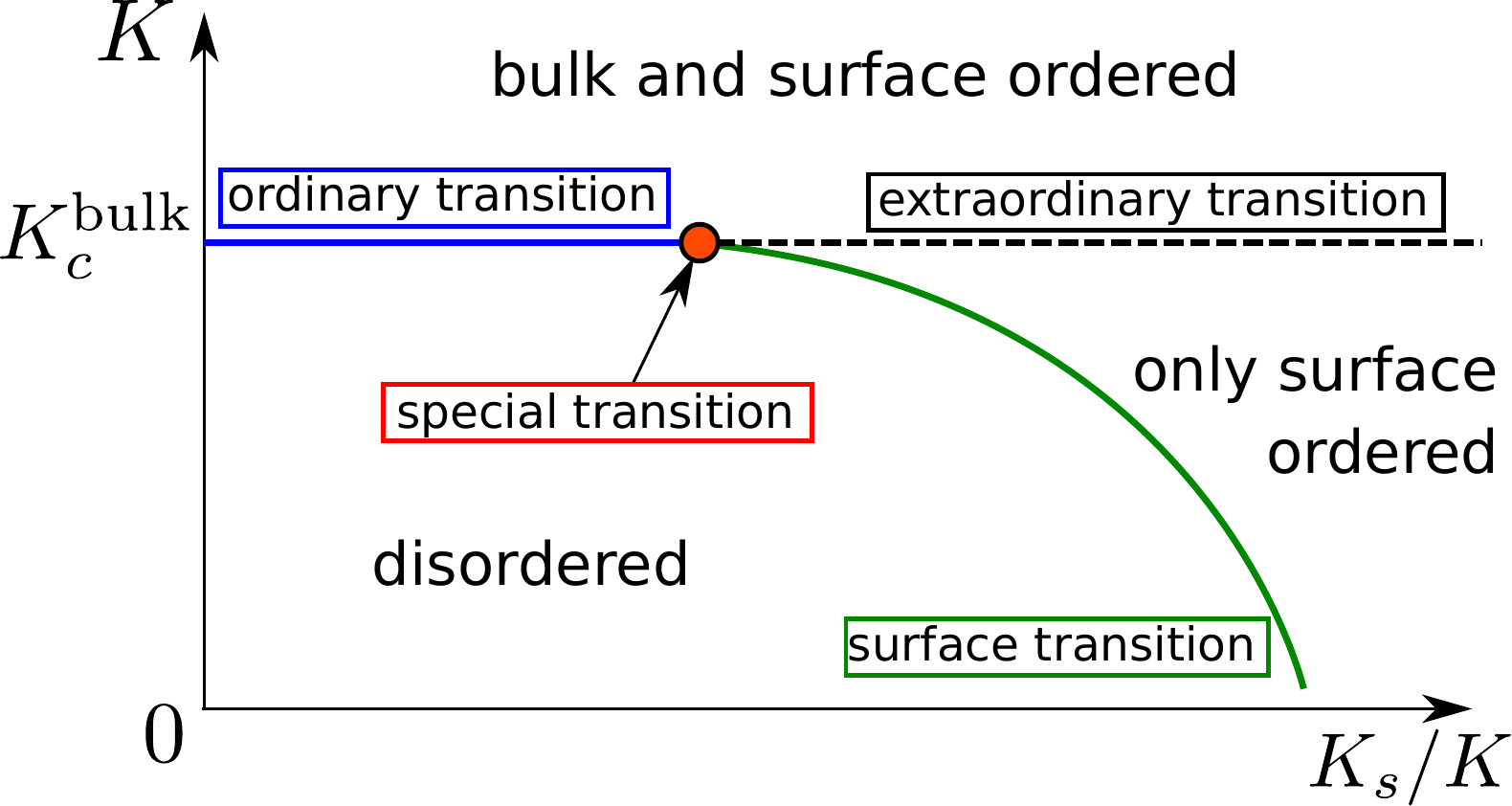}%
\caption{\label{fig:3dIsing}
(Color online) The phase diagram of the three-dimensional classical Ising model with boundaries. $\Kcbulk$ is the bulk critical point of the Ising model in three dimension. There are four different universality classes for the surface critical behavior.}
\end{figure}

Further interesting point of surface critical behaviors is that there are richer universality classes than for the bulk: the surface universality class changes depending on the boundary state at the bulk transition point. A classical example is the Ising model on a semi-infinite cubic lattice with surfaces~\cite{Binder1983_DG}, whose Hamiltonian is
\begin{equation}
  \label{eq:3dIsing}
  \beta\mathcal{H} = -K\sum_{\langle ij\rangle\in\mathrm{bulk}}\sigma_i\sigma_j
  -K_s\sum_{\langle ij\rangle\in\mathrm{surface}}\sigma_i\sigma_j,
\end{equation}
where $\beta$ is an inverse temperature, and the summation of the first term is taken if either of two Ising spins, $\sigma_i$ and $\sigma_j$, belongs to the bulk, while the second summation is done if both of the spins are in the surface. Four surface universality classes may occur by controlling the surface coupling $K_s$ and the bulk one $K$ for ferromagnetic case $K\geq 0$ and $K_s\geq 0$, see Fig.~\ref{fig:3dIsing}. Even when the bulk is disordered in $K<\Kcbulk$, the two-dimensional surface can be ordered by itself if $K_s/K$ is higher than a threshold. This phase transition is called surface transition, whose universality class is believed to be the same as that of two-dimensional Ising model. The transition at $\Kcbulk$ from the disordered phase with disordered surfaces is called ordinary transition. On the other hand, the surface quantities exhibit the different singularity at the transition point from the phase with ordered surfaces, which is called extraordinary transition. Finally there is a special point where the surface transition line and the $K=\Kcbulk$ line merge together, called special transition. These four transitions yield different surface critical exponents in general.

Just as the ordinary CFT successfully explains the critical phenomena on a torus geometry, the description by BCFT is in good agreement with surface critical behaviors. One can consider renormalization group (RG) flows and fixed points for the boundary states just as for various bulk phases, and such boundary fixed points are described by a corresponding BCFT, which tells us the number of the relevant fields, the stability of the fixed points, and the critical exponents of surface quantities. For instance, the two-dimensional Ising BCFT yields three Cardy states corresponding to the three primary fields, $\cardy{\boldsymbol 1}$, $\cardy{\epsilon}$, and $\cardy{\sigma}$, whose physical meanings are the fixed boundary state with plus spins, the one with minus spins, and the disordered (i.e., free) boundary state, respectively~\cite{CARDY1989581}. Therefore, one can calculate the partition function describing, for example, the ordinary transition point by utilizing $\ydrac{\sigma}$ and $\cardy{\sigma}$ for the trace of the partition function. The operator content of this partition function can be calculated as ${\boldsymbol 1}\oplus\epsilon$ by the fusion rule $\sigma\times\sigma = {\boldsymbol 1}+\epsilon$, where $\oplus$ represents the direct sum of two conformal towers. This explains the magnetic critical exponent $\beta_1=1/2$ referred to above, which reflects the relevant field, i.e., the energy operator $\epsilon$ with the conformal dimension $1/2$.

Though the BCFT analysis seems useful to investigate surface critical behaviors as explained above, the relation between critical lattice models and the corresponding BCFTs is in general highly nontrivial. It is always important to confirm the consistency between critical phenomena of lattice models and CFTs, because the emergent conformal invariance on a lattice is just a conjecture.

For some simplest statistical spin models in two dimension at criticality, the Ising, tri-critical Ising, and 3-state Potts model, the boundary fixed points obtained from the Cardy states are theoretically investigated in the framework of BCFT and some integrable models~\cite{CARDY1989581,Chim1996,Affleck1998,Behrend2001}. For real lattice models, while there have been some numerical and exact studies which confirms the validity of the BCFT analyses for the Ising model~\cite{Burkhardt1985,Evenbly2010,Balaska2013}, there is very few study for all the boundary fixed points of the tri-critical Ising and 3-state Potts model. Although some b.c.'s are relatively easy to realize and have been investigated already for some models~\cite{Burkhardt1985,Gehlen1986_2,Gehlen1986,Balbao1987,Lauchli2013,Chepiga2017,Chepiga2019}, the accurate numerical data of higher scaling dimensions are still lacking.

In this paper, we show accurate conformal spectrum for those lattice models with \textit{every} Cardy's boundary state, by which we can conclude the consistency of the BCFT conjectures with the spin models on lattices. As the method to obtain the conformal spectrum, we employ the tensor network renormalization (TNR)~\cite{Evenbly2015}, an accurate numerical method to extract the conformal data of lattice models. Although TNR is originally proposed for the system on a torus geometry, we generalize this algorithm for the system with open boundaries, which can be easily done as implied by Evenbly~\cite{Evenbly2017}. We would like to call this algorithm the \textit{boundary TNR} (BTNR) in this paper.

In the next section Sec.~\ref{sec:BTNR}, we explain the algorithm of BTNR and how to compute the conformal data, after a brief review of tensor renormalization group (TRG) methods. Then, some benchmark results of BTNR using the two-dimensional Ising model are given in Sec.~\ref{sec:benchmark}. In Sec.~\ref{sec:results}, we exhibit the conformal spectrum obtained by the BTNR computation for each boundary fixed points of the tri-critical Ising and 3-state Potts model. Our numerical results are completely consistent with the conjecture of the corresponding BCFTs. Finally, we conclude our work in Sec.~\ref{sec:conclusion}, and in the Appendix explain another method for extracting the conformal spectrum than the one explained in Sec.~\ref{sec:BTNR}.
\section{Boundary Tensor Network Renormalization\label{sec:BTNR}}

In this section, we explain the algorithm of BTNR, and the way of extracting the conformal spectrum. Before discussing the algorithm of BTNR, we begin this section with a brief review of TRG methods. After explaining the predominance of TNR over the ordinary TRG algorithm, we extend TNR so as to perform the RG of boundaries. Finally, we review how to extract the conformal data using the BTNR algorithm, which is already explained in Ref.~\onlinecite{Iino2019}.

\subsection{Overview of TRG methods}

\begin{figure}
\includegraphics[width=8.5cm]{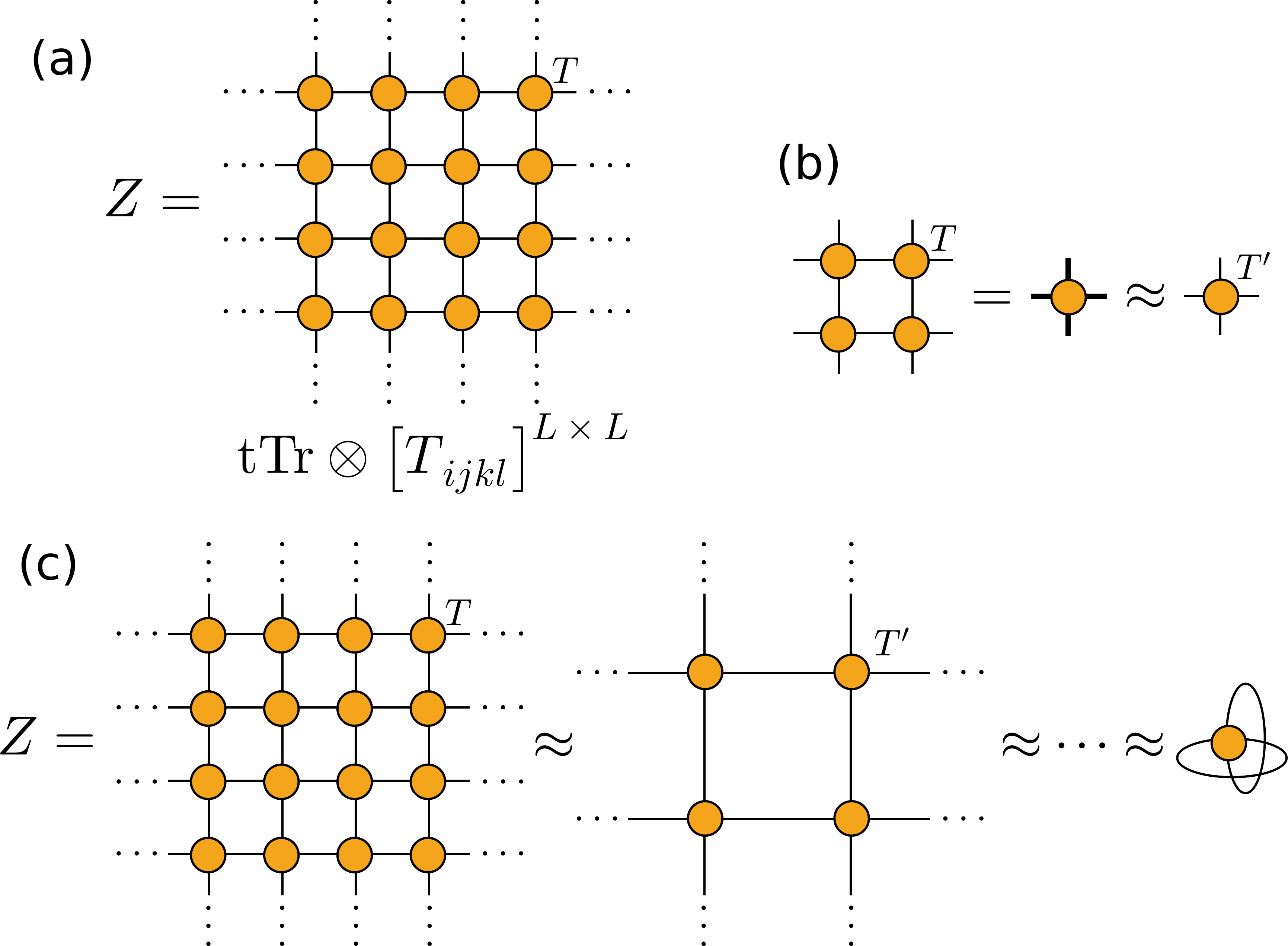}%
\caption{\label{fig:TRG}
(Color online) (a) $Z$ is the contraction of a tensor network constructed by the rank-four tensors $T_{ijkl}$. The tensors are arranged on a square lattice whose system size is $L\times L$. (b) The plaquette network with four tensors $T$ is approximated as a single tensor $T'$. (c) The original network in (a) is replaced by another network consisting of $L/2\times L/2$ tensors $T'$ through the RG procedure in (b). After iterative applications of the RG transformation, one can achieve a network with the countable number of tensors, the contraction of which can be easily taken.}
\end{figure}

TRG is a method to contract tensor networks efficiently~\cite{Levin2007}. Suppose we have a tensor network where rank-four tensors $T_{ijkl}$ are uniformly arranged on a $L\times L$ square lattice, as shown in Fig.~\ref{fig:TRG} (a). For example, the partition function of classical statistical models on a square lattice and the Euclidean path integral of quantum lattice systems on a chain can be represented as this type of tensor networks. Since contracting such a network results in calculation of the partition function, a contraction of tensor networks is significant problem in statistical physics.

TRG methods realize the efficient contraction by employing an approximation based on the real-space RG idea. An example of such an approximation is depicted in Fig.~\ref{fig:TRG} (b), where a tensor network on a plaquette with four tensors $T$ is approximated as a single tensor $T'$. In this RG procedure, it is essential to obtain approximately the renormalized tensor $T'$. If the contraction of the plaquette is taken exactly, the resulting tensor is, as shown in the middle of the equation Fig.~\ref{fig:TRG} (b), a rank-four tensor whose bonds have the squared dimension of the original tensor $T$. This means the RG transformation with the exact contraction leads to an exponential explosion of the bond dimension of a local tensor, which makes it difficult to calculate the contraction of huge networks.

Therefore, the point of the TRG methods is to replace the original tensors efficiently under the condition that the bond dimension of the renormalized tensor $T'$ is limited. To achieve such an efficient approximation, various ways of the RG approximation have been proposed~\cite{Gu2009,Xie2009,Zhao2010,Xie2012,Evenbly2015,Yang2017,Bal2017,Hauru2018,Morita2017,Nakamura2019,Adachi2019,Kadoh2019}. In the next subsection, we introduce an example of a typical TRG algorithm.

As shown in Fig.~\ref{fig:TRG} (c), the use of the RG approximation in Fig.~\ref{fig:TRG} (b) leads to the reduction of the number of local tensors in the original tensor network in Fig.~\ref{fig:TRG} (a): the contraction of the network with $L\times L$ tensors $T$ is replaced by that with $L/2\times L/2$ tensors $T'$. Since repeated application of this RG approximation for the network amounts to the network with the countable number of tensors, one can easily compute the contraction approximately. Figure~\ref{fig:TRG} (c) represents the case where the periodic boundary condition is imposed on the original tensor network, which results in the tensor contraction in the right hand side of the equation.

\subsection{Algorithm of TRG}

\begin{figure}
\includegraphics[width=8.5cm]{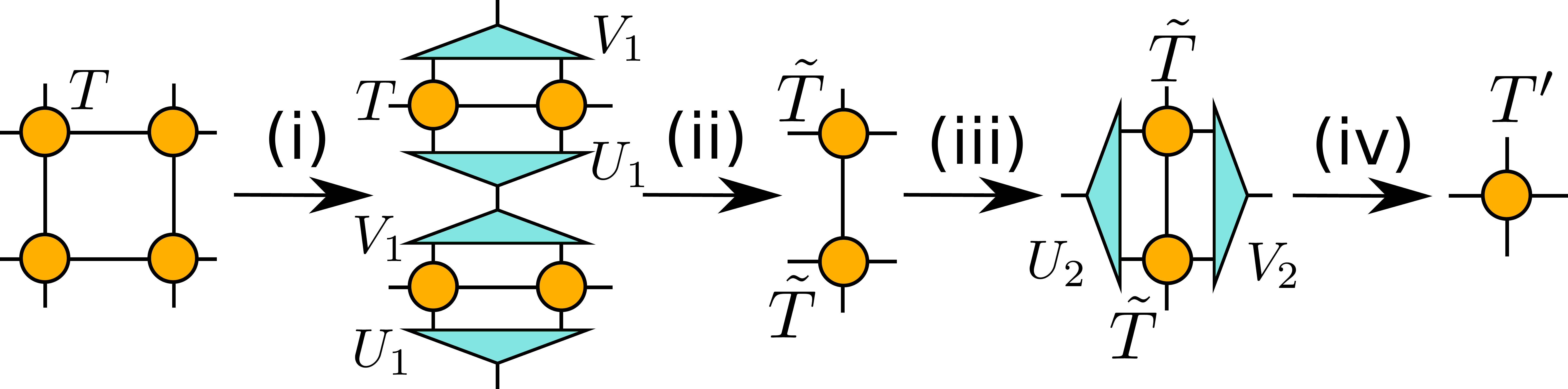}%
\caption{\label{fig:iTRG}
(Color online) One RG step of a typical TRG algorithm. (i) For the horizontal renormalization, the rank-three tensors for the truncation $U_1$ and $V_1$ are inserted into the vertical bonds. (ii) The new tensor $\tilde{T}$ is generated from the contraction of the two horizontally neighbouring tensors $T$ and two tensors $V_1$ and $U_1$. (iii) Then, insert another pair of tensors for the bond truncation $U_2$ and $V_2$ into the horizontal bonds for the vertical renormalization. (iv) After the contraction of the two local tensors $\tilde{T}$, $V_2$ and $U_2$, one can achieve the renormalized tensor $T'$.}
\end{figure}

As we discussed in the last subsection, the core of TRG methods is how to obtain an approximated tensor depicted in Fig.~\ref{fig:TRG} (b). To explain an example of such RG transformation, we briefly introduce a typical TRG algorithm in Fig.~\ref{fig:iTRG}, which is equivalent to the higher order TRG algorithm when the local tensor has an inversion symmetry of the lattice~\cite{Xie2012}. The one RG step can be divided into the horizontal and vertical renormalization, both of which can be completed by the insertion of projectors and the contraction of the local tensors and the projectors.

In Fig.~\ref{fig:iTRG} (i), the pair of the rank-three tensors for the bond truncation $U_1$ and $V_1$ are inserted into every two neighbouring vertical bonds to perform the horizontal renormalization. Such projectors, for instance, can be determined so as to minimize the following cost function:
\begin{equation}
  \mathcal{C}=\left|
  \begin{minipage}{1.5truecm}
      \centering
      \includegraphics[width=1.5truecm,clip]{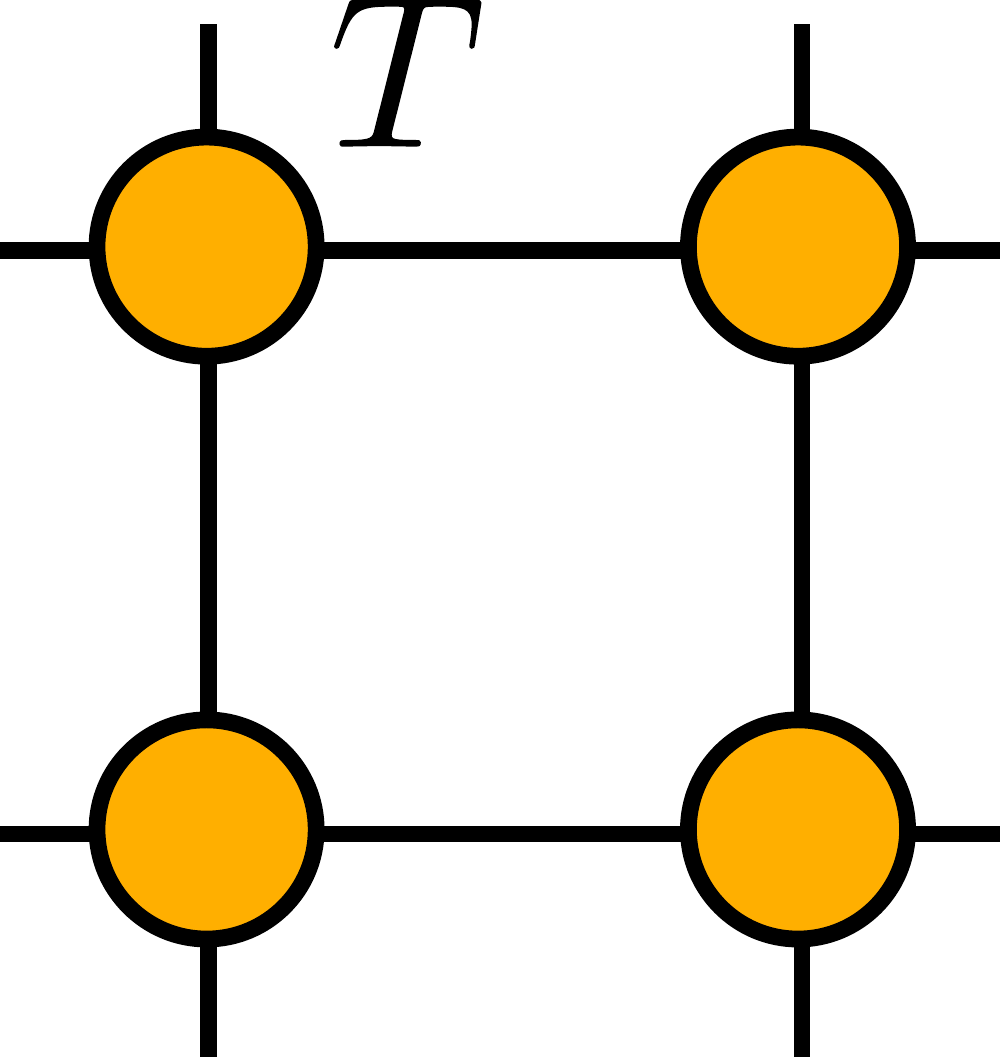}
\end{minipage}
-
\begin{minipage}{1.65truecm}
      \centering
      \includegraphics[width=1.65truecm,clip]{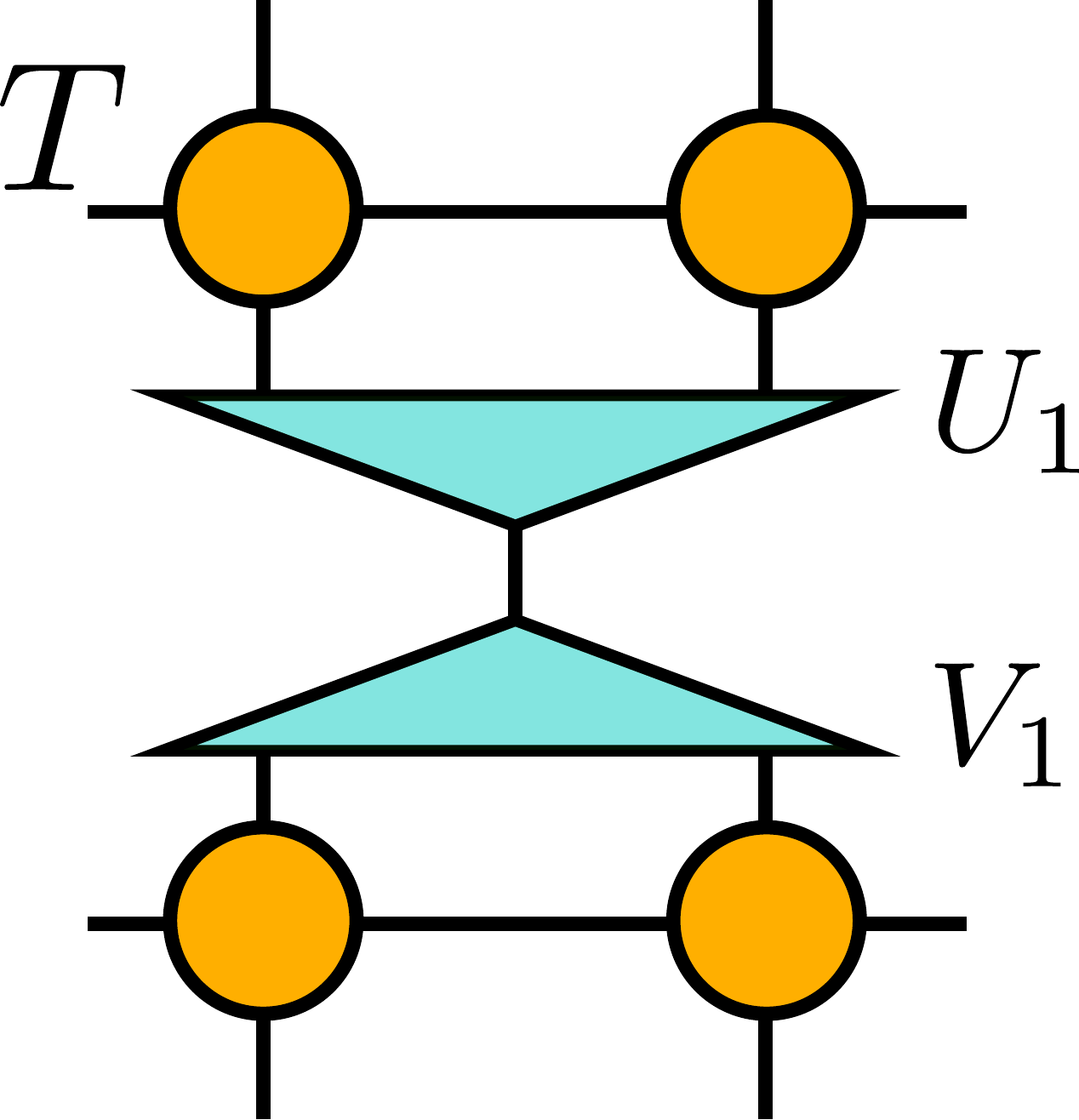}
\end{minipage}
\right|^2.
\label{eq:costfunc}
\end{equation}
It leads to better truncation to make the difference as small as possible between the two networks in \eq{eq:costfunc}. The projector which minimize the cost $\mathcal{C}$ can be obtained by the singular value decomposition (SVD) without iterative optimization, as explained in the appendix A of Ref.~\onlinecite{Iino2019}. After the insertion, a new rank-four tensor $\tilde{T}$ comes from the contraction of the two local tensors $T$ and the two tensors $V_1$ and $U_1$, as in Fig.~\ref{fig:iTRG} (ii). Then another pair of tensors $U_2$ and $V_2$ for the vertical renormalization are immediately inserted into the every neighbouring horizontal bonds as in Fig.~\ref{fig:iTRG} (iii). Note that these projectors can be also computed by minimizing an appropriate cost function similar to \eq{eq:costfunc}. Finally, as in Fig.~\ref{fig:iTRG} (iv), the contraction of the two local tensors $\tilde{T}$, $V_2$ and $U_2$ results in the renormalized tensor $T'$. If we suppose all the bond dimension of the local tensor $T$ is $\chi$, the computational cost of this algorithm scales as $O(\chi^7)$.

\subsection{Fixed point tensor and TNR}

In the paper on the proposal of TRG by Levin and Nave, they pointed out that a tensor converges to some fixed point after enough number of times of RG transformation~\cite{Levin2007}. Gu and Wen discussed the fixed point tensors more precisely and revealed that it is useful to characterize the phases of the system, because the fixed point tensors have the information of the system such as the broken symmetry, the degeneracy, and the conformal data at criticality~\cite{Gu2009}.

However, as Levin and Nave also remarked, the original TRG algorithm has a problem that it cannot eliminate a short correlated loop in a network, which accumulates in the local tensors under successive RG steps. Therefore TRG fails to achieve the correct fixed point tensor, which causes reduction of accuracy especially at criticality. There is detailed discussion on the problem of TRG in Ref.~\onlinecite{Gu2009,Hauru2018}.

\begin{figure}
\includegraphics[width=6.5cm]{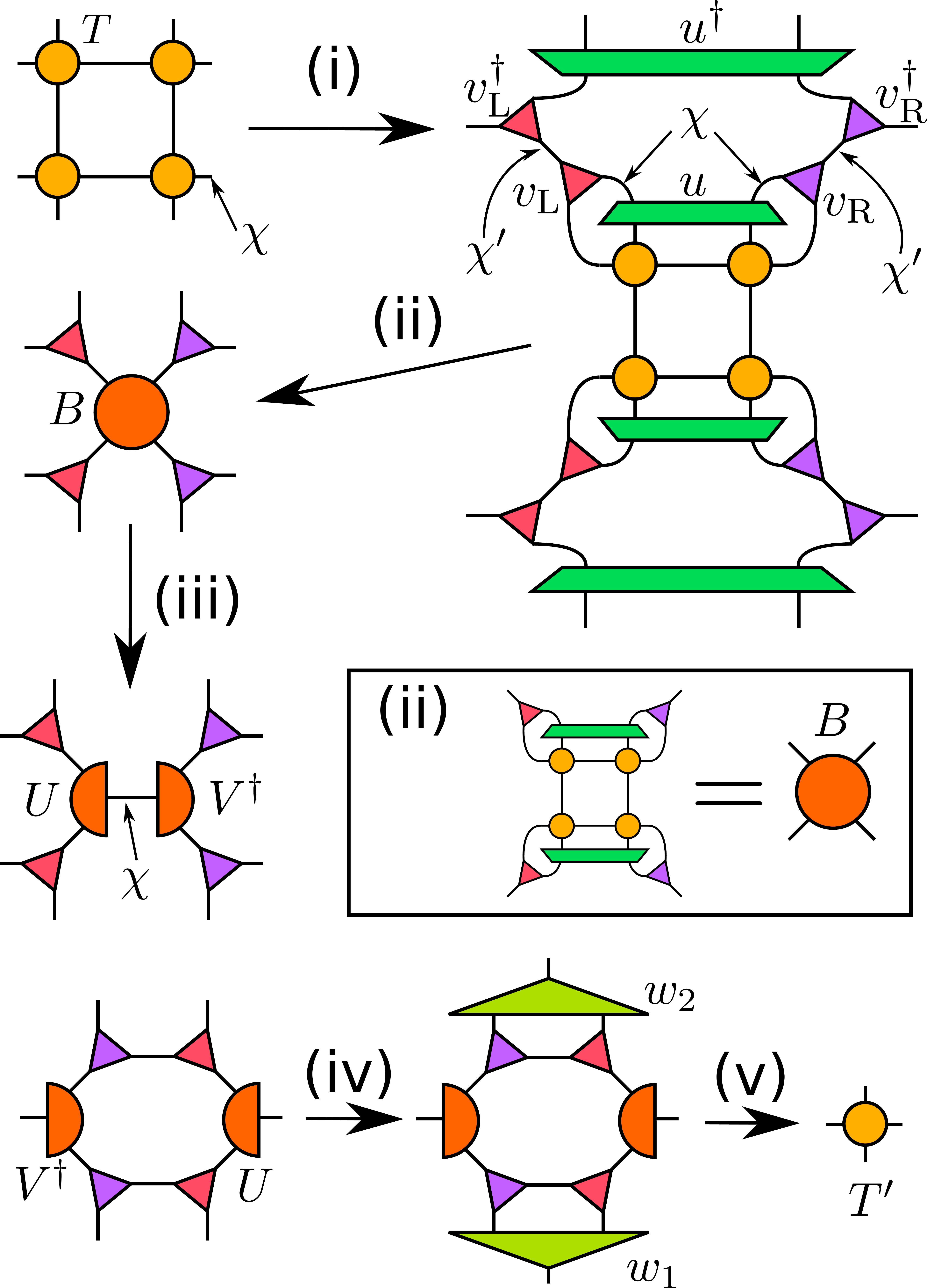}%
\caption{\label{fig:TNR}
(Color online) The RG procedure of the TNR algorithm~\cite{Evenbly2015,Evenbly2017}. The bond dimensions of the original local tensor $T$ are all $\chi$. (i) Insert the isometries $v_{\mathrm{R}}$ and $v_{\mathrm{L}}$, the unitary operator $u$, and the Hermitian conjugates of them $v_{\mathrm{R}}^{\dagger}$, $v_{\mathrm{L}}^{\dagger}$, and $u^{\dagger}$. (ii) After $u^{\dagger}$ at the top and $u$ at the bottom are canceled due to the unitarity of $u$, the new rank-four tensor $B$ is defined as in the figure. (iii) $B$ is decomposed into the two rank-three tensor $U$ and $V^{\dagger}$ with the dimension of the intermediate bond kept below the threshold $\chi$. (iv) To truncate the vertical bonds, insert the two tensors $w_1$ and $w_2$. (v) The contraction gives the renormalized tensor $T'$.}
\end{figure}

TNR algorithm proposed by Evenbly and Vidal resolved this problem of TRG~\cite{Evenbly2015}. TNR successfully yields the approximate fixed-point tensors free from the correlated loops, which leads to much higher accuracy in conformal data than the simple TRG algorithm. The overview of the TNR algorithm is shown in Fig.~\ref{fig:TNR}, and for further details and sophisticated techniques see Ref.~\onlinecite{Evenbly2017}. First of all, and this is the essential point of TNR, the isometries $v_{\mathrm{R}}$ and $v_{\mathrm{L}}$, and the unitary operator $u$, and their Hermitian conjugates $v_{\mathrm{R}}^{\dagger}$, $v_{\mathrm{L}}^{\dagger}$, and $u^{\dagger}$ are inserted as in Fig.~\ref{fig:TNR} (i), which play a roll in disentangling the short-range correlated loop. These isometries and unitary operators are obtained by an iterative optimization method using SVD. Notice that the isometries are $\left(\chi\times\chi\times\chi'\right)$-dimensional tensor as depicted in Fig.~\ref{fig:TNR}, where $\chi$ is the bond dimension of the original local tensor $T$ and $\chi'$ is a squeezed bond dimension. In this study, we adopt $\chi'=\chi/2$. Then, the contraction depicted in Fig.~\ref{fig:TNR} (ii) defines the new rank-four tensor $B$. Notice that the unitary operators in the top and the bottom, $u^{\dagger}$ and $u$, are canceled out because of the unitarity condition: $uu^{\dagger}=u^{\dagger}u={\boldsymbol 1}$. In Fig.~\ref{fig:TNR} (iii), the new tensor $B$ is immediately decomposed into the left tensor $U$ and the right one $V^{\dagger}$, whose intermediate bond is truncated so as not to exceed the threshold $\chi$. Such a decomposition can be performed by, for example, SVD of $B$. The bonds generated by this decomposition will be the horizontal bonds of the renormalized tensor $T'$. After that, the pair of tensors for the truncation $w_1$ and $w_2$ are inserted as in Fig.~\ref{fig:TNR} (iv) for the truncation of the vertical bonds, which can be obtained by the minimization of an appropriate cost function just as \eq{eq:costfunc}. Finally, the contraction of these tensors results in the renormalized tensor $T'$, as shown in Fig.~\ref{fig:TNR} (v). The overall computational cost of the algorithm in Fig.~\ref{fig:TNR} scales as $O(\chi^7)$, although how to reduce the cost to $O(\chi^6)$ is discussed in Ref.~\onlinecite{Evenbly2017}.

\subsection{Algorithm of BTNR}

Now that we have finished the review of TRG and TNR algorithms, let us explain the tensor network method we employ in this study. Since our purpose is to investigate the system with open boundaries, we need to extend the tensor network methods discussed above so as to be applied to the renormalization of not only the bulk but also boundaries. Such generalization of TRG algorithm is proposed by the authors in Ref.~\onlinecite{Iino2019}, which we refer to as boundary tensor renormalization group (BTRG). As shown in Ref.~\onlinecite{Iino2019}, BTRG makes it possible to compute the physical quantities in the boundaries and the boundary conformal spectrum consistent with the corresponding BCFT. In order to obtain more accurate spectrum including the primary or secondary fields with larger conformal dimensions, however, more sophisticated algorithms are required which successfully eliminate the short correlated loop and generate the correct fixed point tensor.

\begin{figure}
\includegraphics[width=7.5cm]{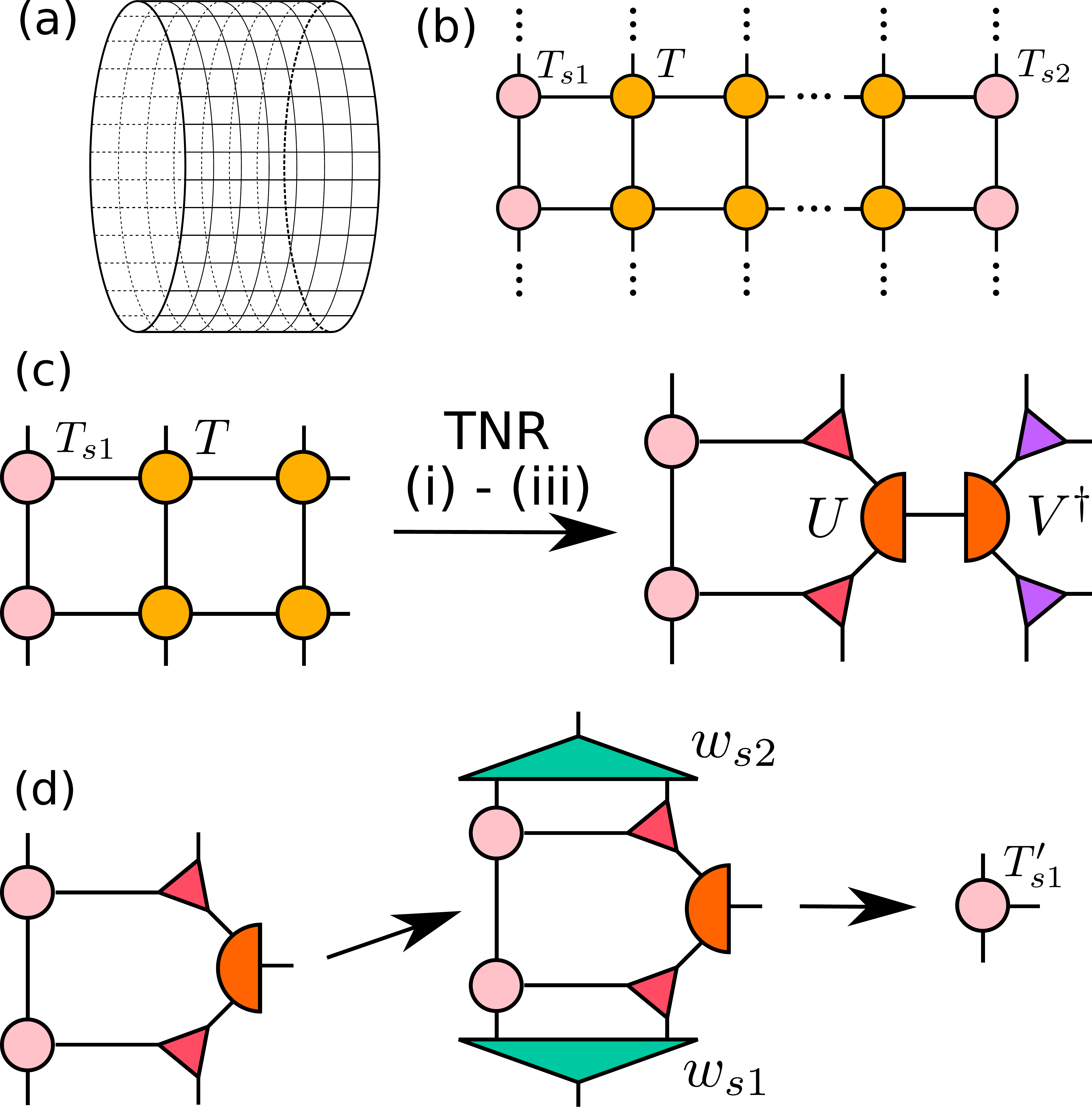}%
\caption{\label{fig:BTNR}
(Color online) (a) A square lattice on a finite cylinder, the geometry to be discussed in this paper. (b) A tensor network on a finite cylinder geometry can be represented using three tensors, the bulk tensor $T$ and the boundary ones $T_{s1}$ and $T_{s2}$. (c) and (d) The RG procedure of BTNR algorithm for the boundary tensor $T_{s1}$. As depicted in (c), the TNR procedures explained in Fig.~\ref{fig:TNR} (i)-(iii) are performed for the plaquette network of the bulk tensors next to the boundary tensors. The decomposition Fig.~\ref{fig:TNR} (iii) generates the $U$ and $V^{\dagger}$, the latter of which will be absorbed into the renormalized bulk tensor $T'$. The leftovers including $U$ and the boundary tensors $T_{s1}$ are renormalized as in (d), where the inserted pair of tensors $w_{s1}$ and $w_{s2}$ and the two boundary tensors are contracted as the renormalized tensor $T_{s1}'$.}
\end{figure}

To accomplish the goal referred to in the last paragraph, we implement the BTNR algorithm, TNR algorithm for the open-boundary systems, the generalization of which is able to be done easily as Evenbly commented in Ref.~\onlinecite{Evenbly2017}. We explain the algorithm of BTNR for a square lattice on a finite cylinder geometry depicted in Fig.~\ref{fig:BTNR} (a), tensor networks on which can be represented by three tensors as in Fig.~\ref{fig:BTNR} (b): the bulk tensor $T$ and the rank-three boundary tensors $T_{s1}$ and $T_{s2}$. Although, for simplicity, we show the renormalization of only the left boundary tensor $T_{s1}$, RG for the other can be performed similarly.

For the RG of the system with open boundary, we have to consider the RG of the bulk and that of the boundary respectively~\cite{Cardy_statmech}, the former of which can be the same as that of RG for the system with the periodic boundary conditions. Therefore, as described in Fig.~\ref{fig:BTNR} (c), we perform RG of the ordinary TNR algorithm for the plaquette network next to the boundary tensors. The procedures shown in Fig.~\ref{fig:TNR} (i), (ii), and (iii) yield the isometries $v_{\mathrm{R}}$, $v_{\mathrm{L}}$, $v_{\mathrm{R}}^{\dagger}$, and $v_{\mathrm{L}}^{\dagger}$, and two rank-three tensors $U$ and $V^{\dagger}$. While the right three tensors, $v_{\mathrm{R}}$, $v_{\mathrm{R}}^{\dagger}$, and $V^{\dagger}$ will contribute to the renormalized bulk tensor $T'$ as in Fig.~\ref{fig:TNR}, the renormalized boundary tensor $T_{s1}'$ is made of the other tensors. After the tensors for the truncation $w_{s1}$ and $w_{s2}$ are inserted, as depicted in Fig.~\ref{fig:BTNR} (d), the renormalized boundary tensor $T_{s1}'$ is generated from the contraction of the $w_{s1}$ and $w_{s2}$, the original boundary tensors $T_{s1}$, $v_{\mathrm{L}}$, $v_{\mathrm{L}}^{\dagger}$, and $U$. The tensors for the truncation are determined again so as to minimize an appropriate cost function just as \eq{eq:costfunc}. Notice that we need not to create any disentangler for the boundary renormalization.

We comment on the comparison with Ref.~\onlinecite{Hauru2016}, where the TNR algorithm for the torus geometry with the line defects is discussed. Since the open boundary can be seen as a line defect, the renormalization of the finite cylinder geometry in Fig.~\ref{fig:BTNR} (a) could be performed similarly to the torus with impurity tensors as in Ref.~\onlinecite{Hauru2016}. This method, however, requires twice per one RG step the most costly optimization procedure in TNR algorithm, because one has to prepare different disentanglers and isometries for the ordinary local tensor and the impurity tensor separately. Since BTNR algorithm demands the optimization only once per one RG step, it can be more economical way of renormalizing the open-boundary system.

Another comment on BTNR is the application to the one-dimensional quantum open chains. Since, similarly to the closed chain~\cite{Evenbly2017}, the Euclidean path integral of the open chain could be represented as the tensor network on a square lattice in a finite cylinder geometry as in Fig.~\ref{fig:BTNR} (b), one can simulate the one-dimensional quantum systems with BTNR by coarse graining this network.

\begin{figure}
\includegraphics[width=8.5cm]{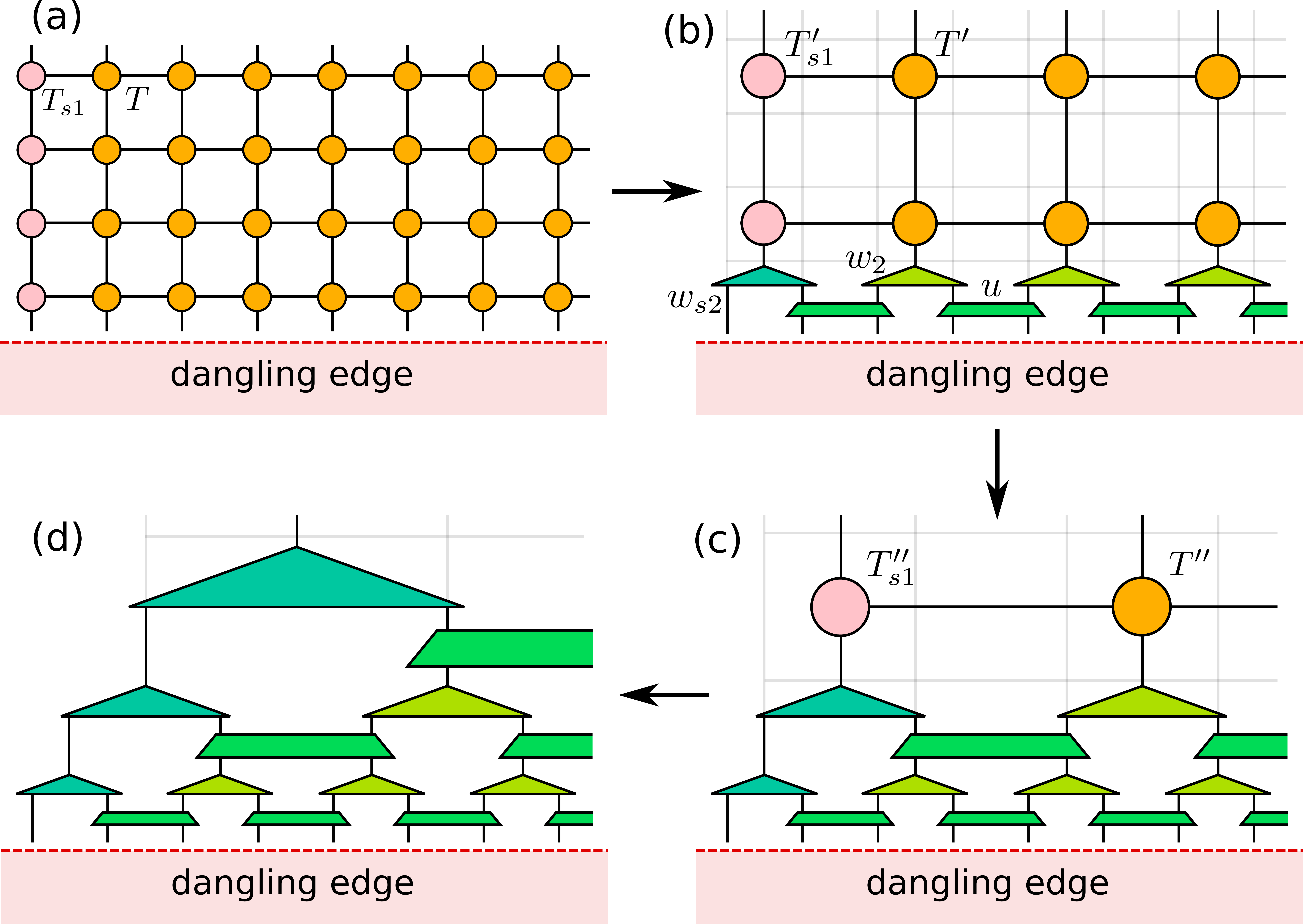}%
\caption{\label{fig:BMERA}
(Color online) The successive RG procedures of BTNR yields the binary BMERA network for the system with the dangling edge and the open boundary. Notice that the open boundary is in the left side of the network, while the dangling edge is the bottom row of the network.}
\end{figure}

By considering the simulation of open quantum chains, we can discuss the relation between BTNR algorithm and the boundary multi-scale entangled renormalization ansatz (boudary MERA, BMERA) network~\cite{Evenbly2010}. As discussed in Ref.~\onlinecite{Evenbly2015_2}, the ordinary TNR algorithm yields the MERA network when one performs the RG of TNR for the Euclidean path integral with the `open boundary', where the bonds in the boundary are dangling and do not connect with anywhere, just as the physical bonds of MERA. Since this `open boundary' is different from the open boundary considered in this paper, we would like to call it the dangling edge in order to avoid confusion. The difference of these two boundaries can be found in Fig.~\ref{fig:BMERA}, where the dangling edge is the bottom row of the tensor networks while the open boundary represents the left side of the networks composed of the rank-three boundary tensors $T_{s1}$. Similarly to the relation between the ordinary TNR and MERA, our BTNR algorithm yields the binary BMERA network for the square lattice with both of the dangling edge and the open boundary, see Fig.~\ref{fig:BMERA}. Therefore, we can derive the BMERA wave-function of a quantum Hamiltonian not only for the ground state but also for thermal states, by applying BTNR for the tensor network of the path integral with a finite temperature, just as in Ref.~\onlinecite{Evenbly2015_2}.

As for the comparison with the binary BMERA as a numerical method for quantum chains, the computational complexity is different: it is $O(\chi^7)$ for our BTNR while $O(\chi^9)$ for the binary BMERA. This suggests that for the simulation of open quantum chains, BTNR can allow us to avoid the numerically expensive optimization of BMERA, although detailed comparison is out of the scope of the present work.

\subsection{Computation of the conformal spectrum}

In this subsection, we briefly explain how to calculate the conformal spectrum from a coarse grained tensor network at criticality, which is discussed in detail in Ref.~\onlinecite{Iino2019}.

First of all, we calculate the scale-invariant boundary tensors $T_{s1\mathrm{inv}}$ and $T_{s2\mathrm{inv}}$ by dividing the local boundary tensors $T_{s1}$, and $T_{s2}$ by the appropriate factor, for the detail of which see the appendix B of Ref.~\onlinecite{Iino2019}. The transfer matrix made from the scale-invariant tensors can be related to quantities appearing in BCFT through the universal term of the partition function, as
\begin{equation}
  Z_{\mathrm{BCFT}}=\tTr\left[
  \begin{minipage}{1.1truecm}
      \centering
      \includegraphics[width=1.1truecm,clip]{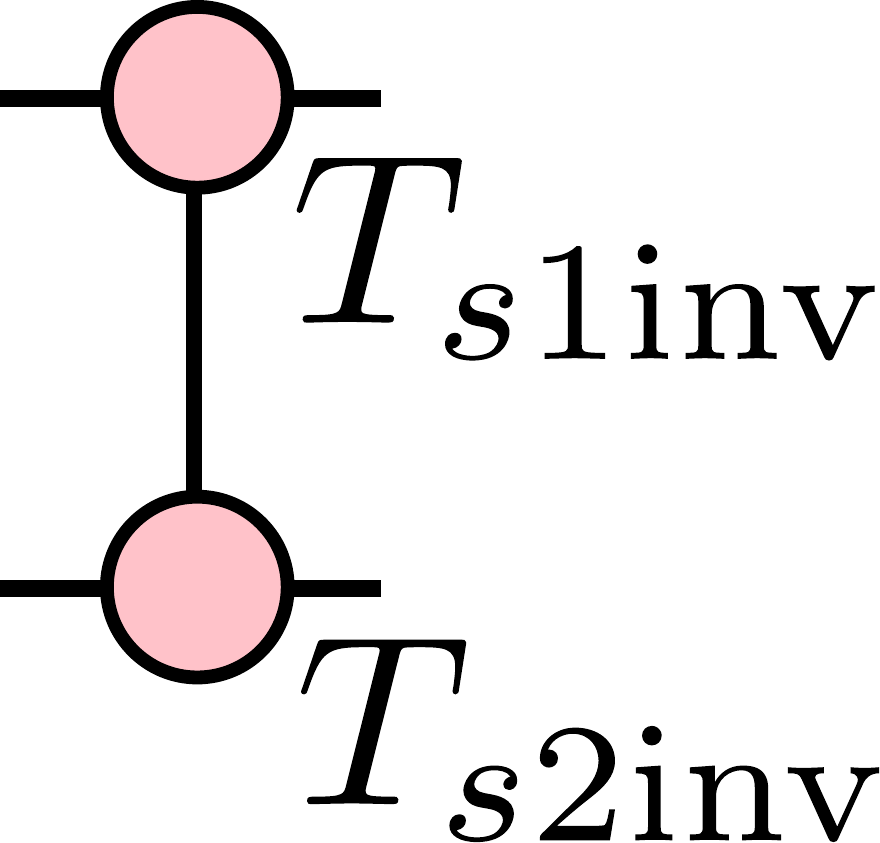}
  \end{minipage}
  \right]^M
  =\mathrm{Tr}\exp\left[-\frac{M}{2}\pi\left(\hat{L}_0-\frac{c}{24}\right)\right],
\label{eq:BCFTZ}
\end{equation}
where $M$ is the circumference of the cylinder, $\hat{L}_0$ is the Virasoro operator, and $c$ is the central charge. Thanks to this equation, the eigenvalue spectrum of the transfer matrix from the tensor network, which is labeled by an integer $n$ in descending order as $\{\lambda_n\}$, give the conformal dimensions $\{h_n\}$ as
\begin{equation}
  \ln\lambda_n = -\frac{\pi}{2}\left(h_n-\frac{c}{24}\right).
  \label{eq:spectrum-formula}
\end{equation}
Notice that one cannot completely determine the unknown quantities in the right hand side, $c$ and $\{h_n\}$, only from the eigenvalue spectrum $\{\lambda_n\}$, since the number of the unknown quantities is one more than that of the equations. Therefore, it is necessary to determine the value of one of the unknown quantities without \eq{eq:spectrum-formula}. In studying BCFT, its ordinary CFT is often known beforehand, in which case one can utilize the known central charge $c$ in \eq{eq:spectrum-formula}. In this paper, we study the boundary fixed points where both of two edges in the finite cylinder are the same boundary states. When we investigate such boundary fixed points, the identity ${\boldsymbol 1}$ would be in the operator content, which allows us to assume $h_0=0$ for the unitary CFTs. In this study we are making use of this information on $h_0$, and obtain the central charge and the other conformal dimensions.

Finally, we would like to comment that the conformal spectrum can also be obtained by employing the super operator for the boundary tensors, which is explained in the Appendix. However, since this method requires costly computational resources, as the way of extracting conformal data we adopt the technique explained in this subsection.
%

\section{Benchmark Results of BTNR\label{sec:benchmark}}

In this section, we show the results of the application of BTNR for the two-dimensional Ising model,
\begin{eqnarray}
  \beta\mathcal{H}=-\Kcbulk\sum_{\langle ij\rangle\in\mathrm{bulk}}\delta_{\sigma_i\sigma_j}
  &-&K_s\sum_{\langle ij\rangle\in\mathrm{surface}}\delta_{\sigma_i\sigma_j}\nn
  &-&h_s\sum_{i\in\mathrm{surface}}\sigma_i.
  \label{eq:potts-hamiltonian}
\end{eqnarray}
where $\sigma$ takes $+1$ or $-1$ and $\Kcbulk$ is the exact critical point of the bulk transition at $\Kcbulk=\ln\left(1+\sqrt{2}\right)$~\cite{Ising1925}. We adopt the $Z_2$ symmetric tensor when simulating $h_s=0$~\cite{Singh2011}. The purpose of this section is especially to demonstrate that the BTNR algorithm yields the correct scale-invariant fixed point tensor at the critical point, which cannot be obtained by the simple BTRG algorithm. To show the evidence of the correct fixed point tensor, we compute entropies for the eigenvalue spectrum of a coarse grained transfer matrix and the conformal spectrum, which are stable against the increment of RG step, while BTRG produces monotonically increasing entropy and much narrower steady region. This stability is a hallmark of generating the correct fixed point structure.

\subsection{Entropy of the eigenvalue spectrum of the transfer matrix}

First of all, we check entropies of the eigenvalue spectrum for the transfer matrix constructed from the boundary local tensors, which can be computed using the following matrix and quantities:
\begin{eqnarray}
  \label{eq:entropy-Gamma}
  \Gamma&=&
  \begin{minipage}{1.2truecm}
      \centering
      \includegraphics[width=1.2truecm,clip]{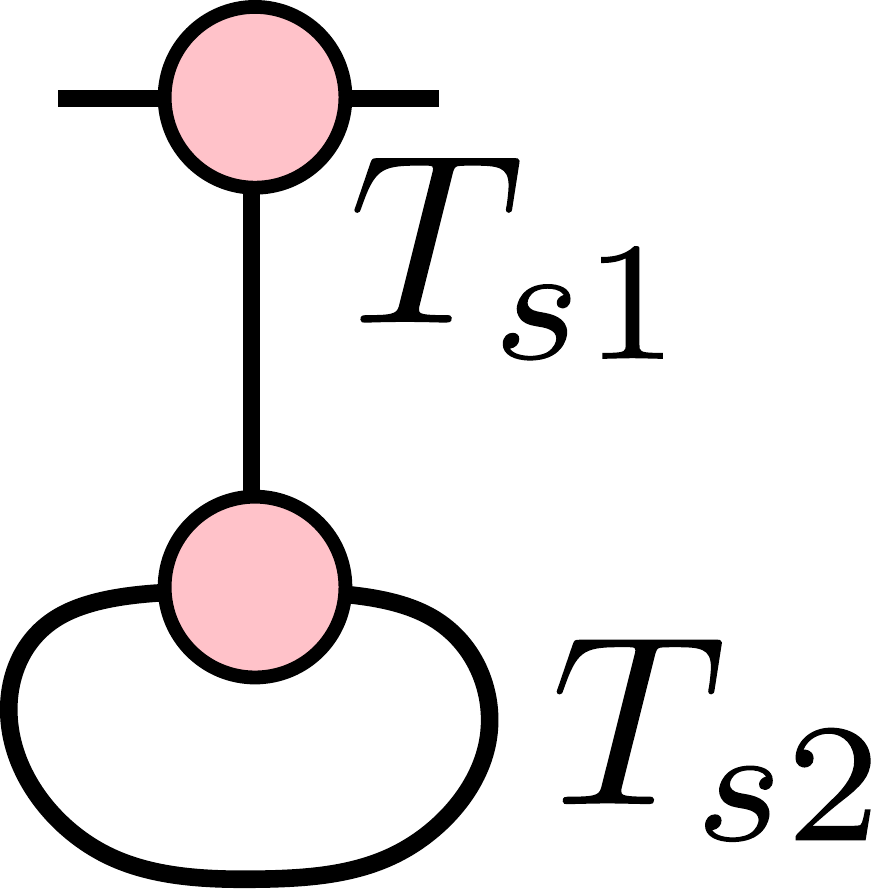}
  \end{minipage},\\
  \Theta_1 &=&\tTr\left(\Gamma\right)=
  \begin{minipage}{0.7truecm}
      \centering
      \includegraphics[width=0.7truecm,clip]{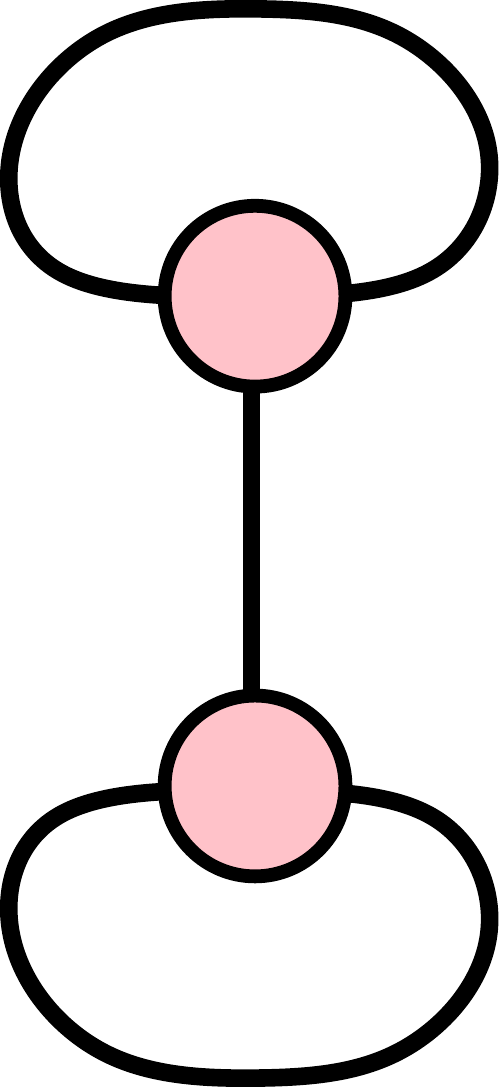}
  \end{minipage},\\
  \Theta_2&=&\tTr\left(\Gamma^2\right)=
  \begin{minipage}{1.7truecm}
      \centering
      \includegraphics[width=1.7truecm,clip]{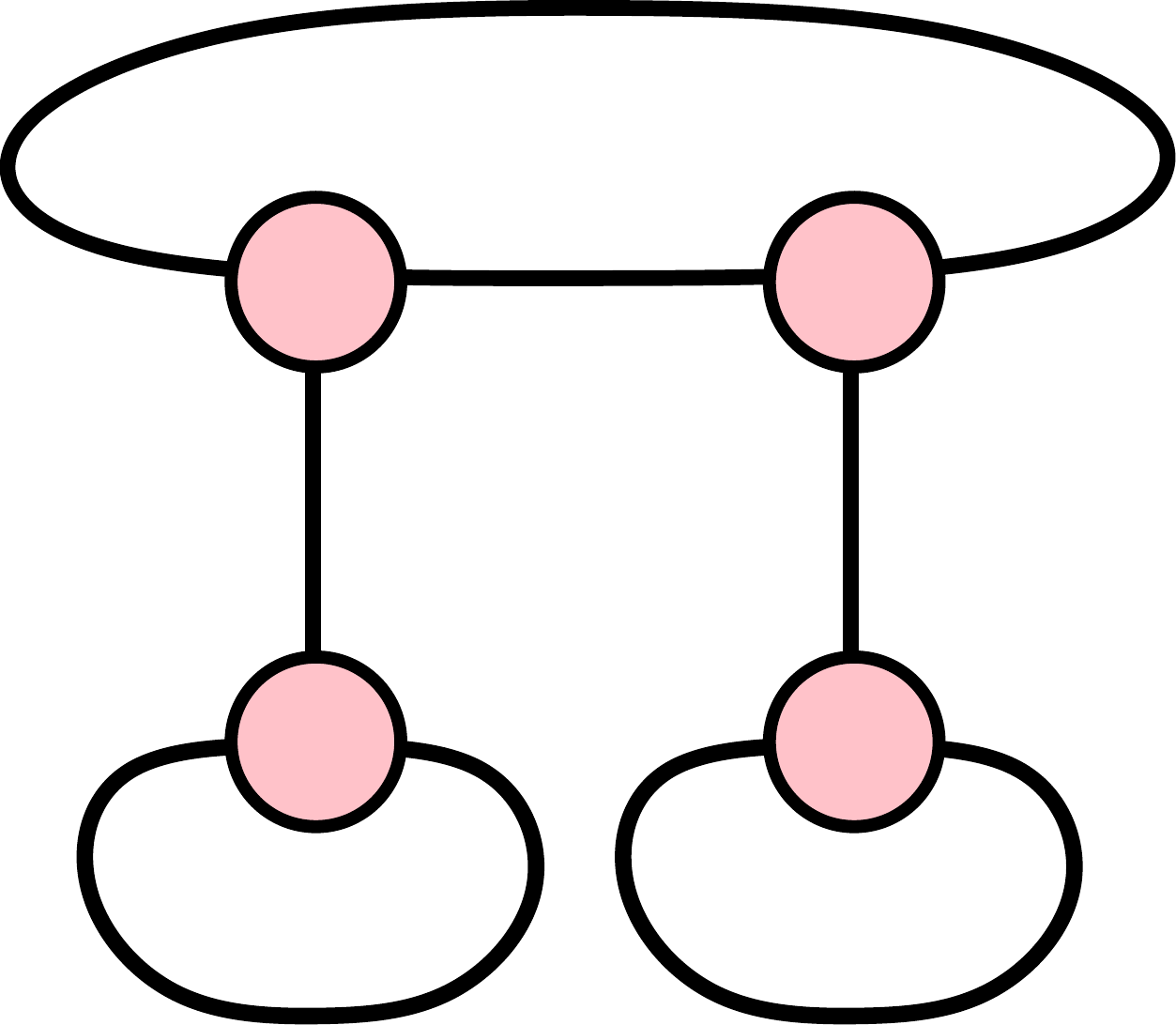}
  \end{minipage}.
\end{eqnarray}
We define $\{\Lambda_n\}$ as the eigenvalues of the matrix $\Gamma/\Theta_1$, from which the von Neumann entropy $S$ can be calculated as
\begin{equation}
  S(\{\Lambda_n\}) = -\sum_{n}\Lambda_n\ln\Lambda_n,
  \label{eq:vN}
\end{equation}
while the Renyi entropy whose degree is two can be defined as
\begin{equation}
  R_2(\{\Lambda_n\}) = -\ln\sum_{n}{\Lambda_n}^2 = -\ln\frac{\Theta_2}{{\Theta_1}^2}.
  \label{eq:R}
\end{equation}
These definitions are the naturally generalized ones discussed in the original paper on TNR~\cite{Evenbly2015}, which are defined so as to be invariant under an arbitrary gauge transformation for the bonds of the local tensors. Notice that these quantities in general are not identical to the entanglement entropy between two subsystems. They are simply measures of the correlation mediated by the tensors, which are in many cases hard to relate to the entanglement of correlation between two well-defined subsystems.

\begin{figure}
\includegraphics[width=6cm]{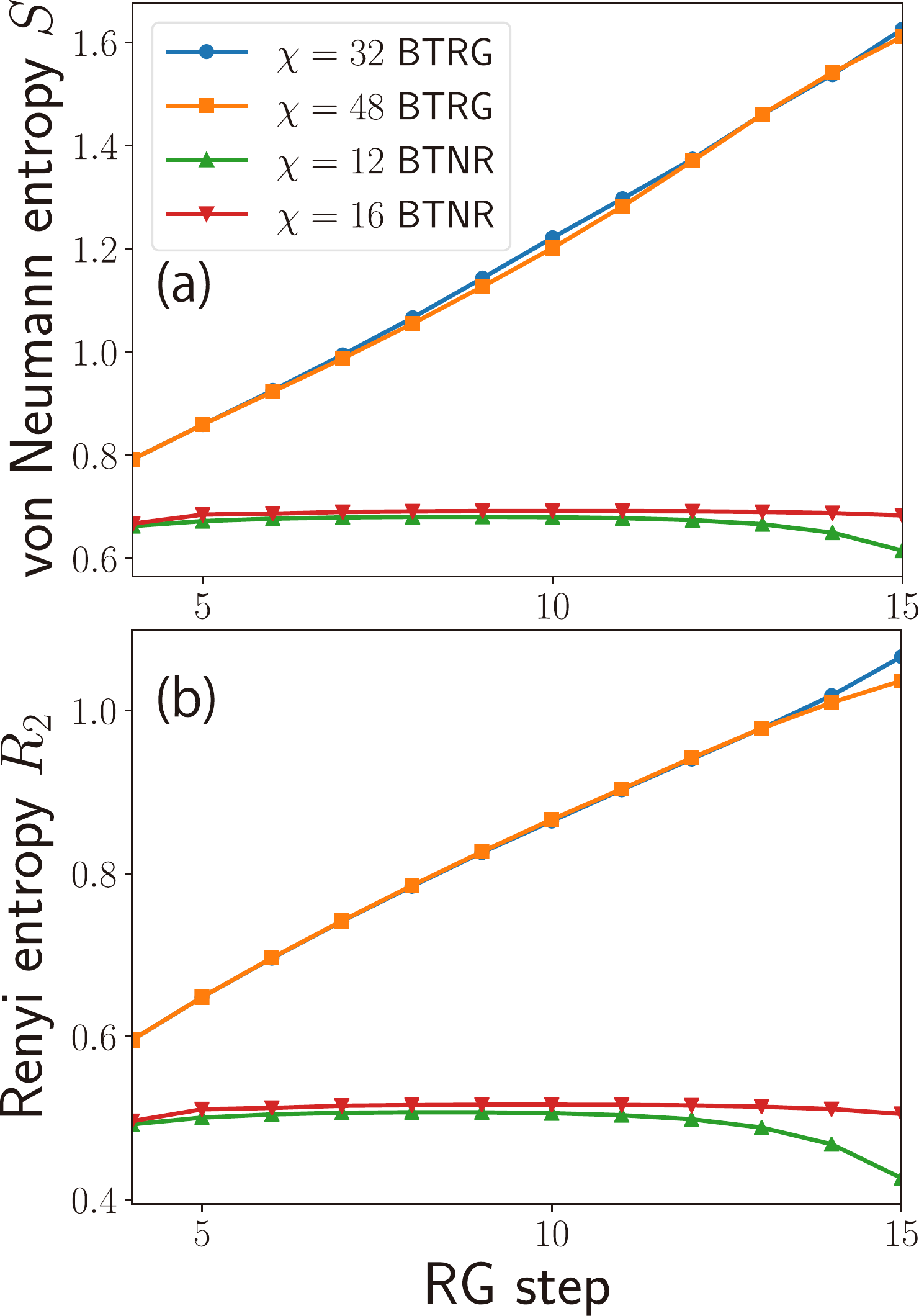}%
\caption{\label{fig:entropy}
(Color online) (a) The von Neumann entropy computed by \eq{eq:vN} and (b) the Renyi entropy computed by \eq{eq:R} at each RG step, for BTRG and BTNR algorithms with the several bond dimensions.}
\end{figure}

The numerical results are shown in Fig.~\ref{fig:entropy} for several bond dimensions with the boundary coupling $K_s=0$ and the external field $h_s=0$, which is equivalent to impose the free boundary conditions on the cylinder. For both of the von Neumann entropy and the Renyi one, the flows of the entropies are flat for BTNR, while those obtained by BTRG grow up almost linearly as a function of the RG steps, which results from the accumulation of the short entangled loops. The stable flows of the entropies suggest that BTNR algorithm removes the short correlated loops and achieves the correct fixed point tensor. Though not shown in the figure, the computed entropies with $K_s=\infty$ and $h_s=0$, which amounts to the fixed boundary condition, have the same tendency as in Fig.~\ref{fig:entropy}.

\subsection{Flow of the scaling dimensions}

\begin{figure*}
\includegraphics[width=15cm]{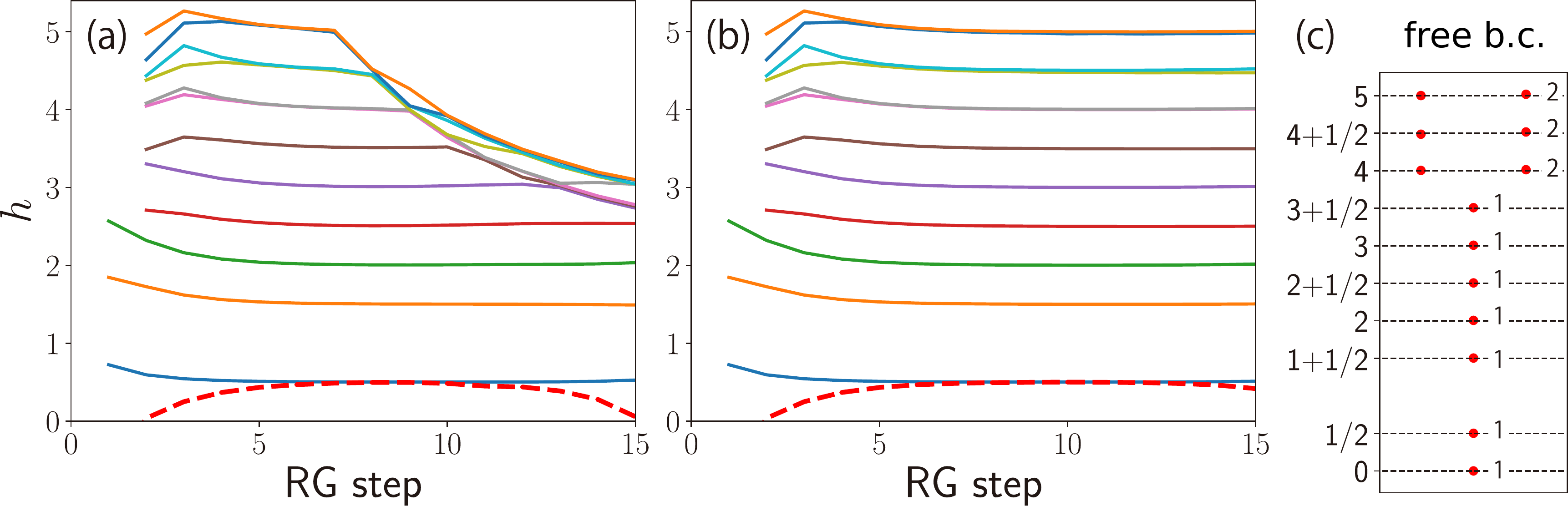}%
\caption{\label{fig:q2_free}
(Color online) (a) and (b) The conformal spectrum (the solid lines) and the central charge (the red dashed line) computed for the Ising model with the free boundary conditions, $K_s=0$ and $h_s=0$. The results are computed (a) by BTRG with $\chi=48$ and (b) BTNR with $\chi=36$. (c) The conformal spectra obtained by BTNR with $\chi=28$ at the eighth RG step, compared with the exact Ising BCFT results. The dashed line and the small figure near the plots represent the exact values of the conformal dimensions and the exact degeneracies, respectively.}
\end{figure*}

In this subsection, we show the numerically obtained boundary conformal spectrum for \eq{eq:potts-hamiltonian}, which is calculated by \eq{eq:spectrum-formula} with the assumption $h_0=0$. We plot the smallest scaling dimensions whose exact values are less than or equal to five and the central charge computed at each RG step under $K_s=0$ and $h_s=0$ in Fig.~\ref{fig:q2_free}. BTRG gives the unstable flow of the dimensions as in Fig.~\ref{fig:q2_free} (a), where the larger dimensions collapse after the several RG steps due to the accumulation of the short entangled loops. As in Fig.~\ref{fig:q2_free} (b), on the other hand, the resulting flow of the conformal spectrum by BTNR is highly stable.
\begin{table*}[htb]
\caption{The extracted scaling dimensions for the free b.c. by BTNR with $\chi=28$ at the eighth RG step, which are plotted in Fig.~\ref{fig:q2_free} (c). The obtained central charge is $c=0.498$, which is consistent with the exact value $c=0.5$.\label{tab:q2_free}}
\begin{ruledtabular}
\begin{tabular}{cccccccccccccccccc}
  exact & 0.5 & 1.5 & 2 & 2.5 & 3 & 3.5 & 4 & 4 & 4.5 & 4.5 & 5 & 5 & 5.5 & 5.5 & 6 & 6 & 6 \\
  BTNR & 0.501 & 1.504 & 2.006 & 2.506 & 3.008 & 3.508 & 4.005 & 4.014 & 4.487 & 4.517 & 5.002 & 5.020 & 5.472 & 5.521 & 5.923 & 5.984 & 6.024\\
\end{tabular}
\end{ruledtabular}
\end{table*}

As referred to in Sec.~\ref{sec:intro}, the operator content for the free boundary fixed point is calculated as ${\boldsymbol 1}\oplus\epsilon$, which is completely consistent with the plots in Fig.~\ref{fig:q2_free} (c) and Tab.~\ref{tab:q2_free}, the spectra at the eighth RG step from BTNR with $\chi=28$.

\subsection{The conformal spectrum for the other fixed points\label{subsec:IsingFP}}

\begin{figure}
\includegraphics[width=8.5cm]{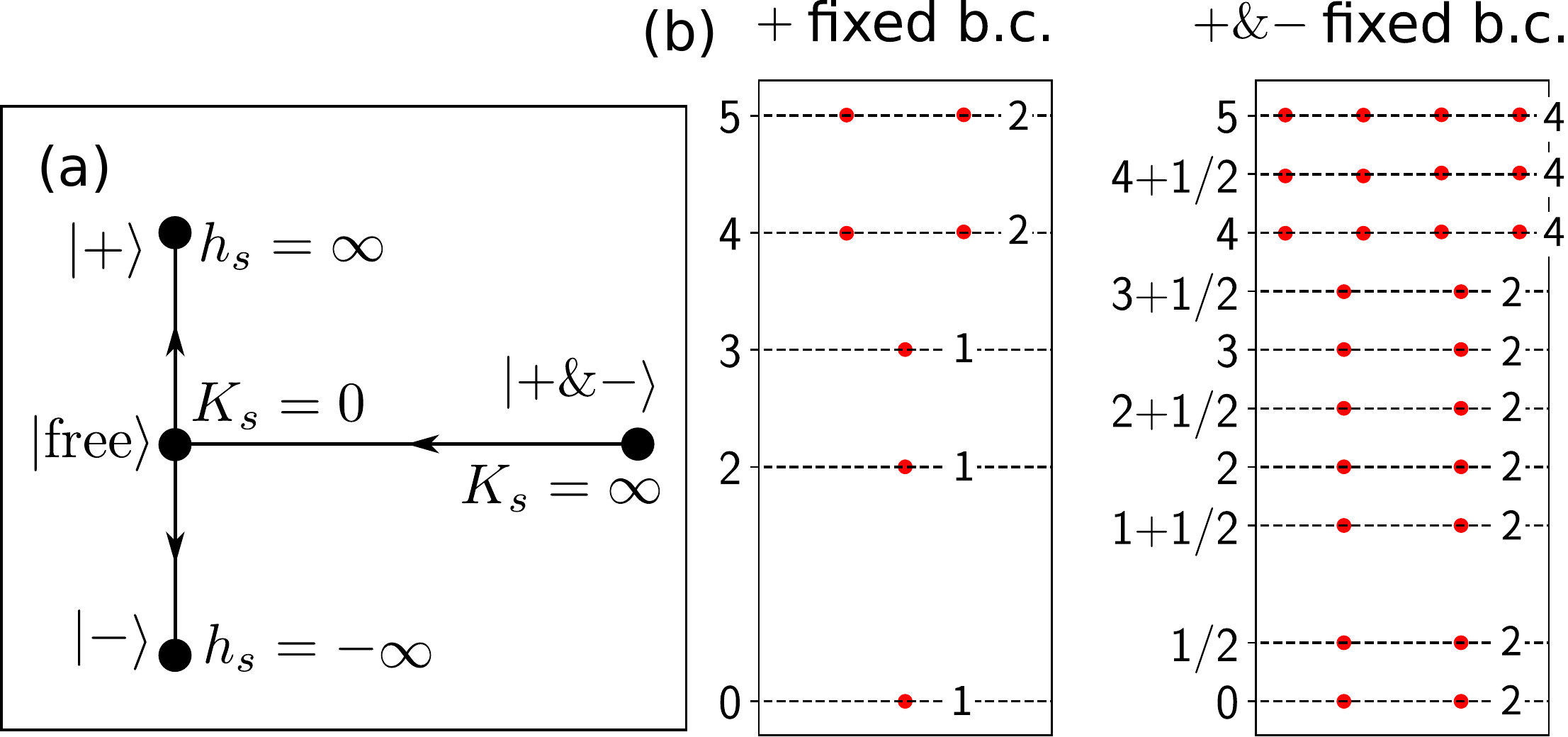}%
\caption{\label{fig:q2_fixed}
(Color online) (a) The boundary phase diagram of the Ising model in two dimension. $\Ket{+\&-}$ is defined as the spontaneously ordered boundary state, $\Ket{+}+\Ket{-}$. (b) The conformal spectrum for the fixed point of the $+$ fixed b.c. at $h_s=\infty$ and $K_s=0$ and the $+\&-$ fixed b.c. at $K_s=\infty$ and $h_s=0$, both of which are computed by BTNR with $\chi=28$ at the eighth RG step.}
\end{figure}

In this subsection, we exhibit the scaling dimension spectrum of the two-dimensional Ising model for the other fixed points than the free boundary. The phase diagram of the boundary states for \eq{eq:potts-hamiltonian} is described in Fig.~\ref{fig:q2_fixed} (a). When both of the surface coupling and external field are zero, $K_s=0$ and $h_s=0$, the surface is disordered, which means the free boundary state $\Ket{\mathrm{free}}$. This fixed point is unstable for the direction of a finite surface external field $h_s$: an infinitesimal $h_s$ induces the $Z_2$ symmetry broken states $\Ket{+}$ or $\Ket{-}$ depending on the sign of $h_s$. The boundary states $\Ket{+}$ and $\Ket{-}$ represent the fixed boundary condition with the $+$ spin and $-$ spin, respectively. Though the fixed point of the free b.c. is stable against the finite surface coupling, the infinite $K_s$ results in the ordered boundary state $\Ket{+\&-}\equiv\Ket{+}+\Ket{-}$, where the two symmetry-broken fixed boundary states are degenerated. Note that this boundary states can be dealt with the $Z_2$ symmetric tensor.

Figure~\ref{fig:q2_fixed} (b) and (c) show the conformal spectrum for the boundary fixed point of the $+$ and $+\&-$ fixed b.c.'s, obtained by BTNR algorithm at the eighth RG step with $\chi=28$. While the operator content for the fixed point of the $+$ fixed b.c. is only ${\boldsymbol 1}$, that for the other can be also calculated by the fusion rules as follows~\cite{Affleck2000}:
\begin{eqnarray}
  \left[{\boldsymbol 1}+\epsilon\right]\times\left[{\boldsymbol 1}+\epsilon\right]
  &=&\left[{\boldsymbol 1}\times{\boldsymbol 1}\right]+\left[{\boldsymbol 1}\times\epsilon\right]
  +\left[\epsilon\times{\boldsymbol 1}\right]+\left[\epsilon\times\epsilon\right]\nn
  &=&2\left[{\boldsymbol 1}+\epsilon\right],
\end{eqnarray}
because of $\Ket{+}=\cardy{\boldsymbol 1}$ and $\Ket{-}=\cardy{\epsilon}$, from which $\Ket{+\&-}$ can be obtained as the sum of the Cardy states $\cardy{\boldsymbol 1}+\cardy{\epsilon}$. The computed spectrum are consistent with the analytical results from Ising BCFT. Notice that the conformal spectra for the fixed point of the $-$ fixed b.c. at $h_s=-\infty$ is trivially the same as that for the $+$ fixed one.
%

\section{Main results\label{sec:results}}

In this section, we show the BTNR computation of the boundary conformal spectrum for the tri-crtical Ising and 3-state Potts model. The Cardy's condition yields 6 boundary states for the tri-critical Ising model and 8 boundary states for 3-state Potts model, for each of which we numerically obtain the spectrum consistent with the conjecture from the corresponding BCFTs.

\subsection{Tri-critical Ising model}

We realize the tri-critical Ising universality class using the Blume-Capel model whose Hamiltonian is
\begin{eqnarray}
  \beta\mathcal{H}=&-&K_{\mathrm{c}}^{\mathrm{bulk}}\sum_{\langle ij\rangle\in\mathrm{bulk}}\sigma_i\sigma_j+D_{\mathrm{c}}^{\mathrm{bulk}}\sum_{i}{\sigma_i}^2\nn
  &-&K_s\sum_{\langle ij\rangle\in\mathrm{surface}}\sigma_i\sigma_j-h_s\sum_{i\in\mathrm{surface}}\sigma_i,
  \label{eq:BC-hamiltonian}
\end{eqnarray}
with $\sigma=0$ or $\pm 1$~\cite{Blume1966,Capel1966}. The bulk coupling and the chemical potential are tuned at the bulk tri-critical point $\Kcbulk=1.6431758$ and $D_{\mathrm{c}}^{\mathrm{bulk}}=3.2301797$~\cite{Deng2004,Qian2005}.

The Cardy states of this model are listed in Tab.~\ref{tab:TCI} corresponding to the six primary fields of the unitary minimal CFT $\mathcal{M}_{5,4}$: ${\boldsymbol 1}(0)$, $\epsilon(1/10)$, $\epsilon'(3/5)$, $\epsilon''(3/2)$, $\sigma(3/80)$, and $\sigma'(7/16)$, where the values in the parentheses represent the conformal dimensions of the primary fields. $\Ket{+\&-}$ is also defined just as in the analysis of the Ising model in Sec.~\ref{subsec:IsingFP} in addition to the six Cardy states. Remember that the operator content of the boundary fixed point can be derived from the fusion rule of the primary field corresponding to the Cardy state. The operator content for the $+\&-$ boundary fixed point can be calculated as~\cite{Affleck2000}
\begin{equation}
  \left[{\boldsymbol 1}+\epsilon''\right]\times\left[{\boldsymbol 1}+\epsilon''\right]=
  2\left[{\boldsymbol 1}+\epsilon''\right],
\end{equation}
which means that the ${\boldsymbol 1}$ and $\epsilon''$ occur twice in the conformal spectrum.
\begin{table}[htb]
\caption{The Cardy states of the tri-critical Ising model and the operator content of the corresponding boundary fixed points. $(K_s,h_s)$ represents the location in the parameter space where the numerical computation has been performed.}
\label{tab:TCI}
\begin{ruledtabular}
  \begin{tabular}{ccc}
  Cardy state & operator content & $(K_s,h_s)$\\\hline
   $\Ket{+}\equiv\cardy{\boldsymbol 1}$ & ${\boldsymbol 1}$ & $(0,+\infty)$ \\
   $\Ket{-}\equiv\cardyw{\epsilon''}$ & ${\boldsymbol 1}$ & \\\hline
   $\Ket{0+}\equiv\cardy{\epsilon}$ & ${\boldsymbol 1}\oplus\epsilon'$ & $(\Kcbulk,0.67721)$ \\
   $\Ket{0-}\equiv\cardyw{\epsilon'}$ & ${\boldsymbol 1}\oplus\epsilon'$ & \\\hline
   $\Ket{0}\equiv\cardyw{\sigma'}$ & ${\boldsymbol 1}\oplus\epsilon''$ & $(0,0)$\\\hline
  $\Ket{\mathrm{d}}\equiv\cardy{\sigma}$ &
  ${\boldsymbol 1}\oplus\epsilon\oplus\epsilon'\oplus\epsilon''$ & $(1.56623\times\Kcbulk,0)$ \\\hline
  $\Ket{+\&-}\equiv\Ket{+}+\Ket{-}$ & $2\left({\boldsymbol 1}\oplus\epsilon''\right)$ & $(\infty,0)$ \\
 \end{tabular}
\end{ruledtabular}
\end{table}
%

\begin{figure*}
\includegraphics[width=18cm]{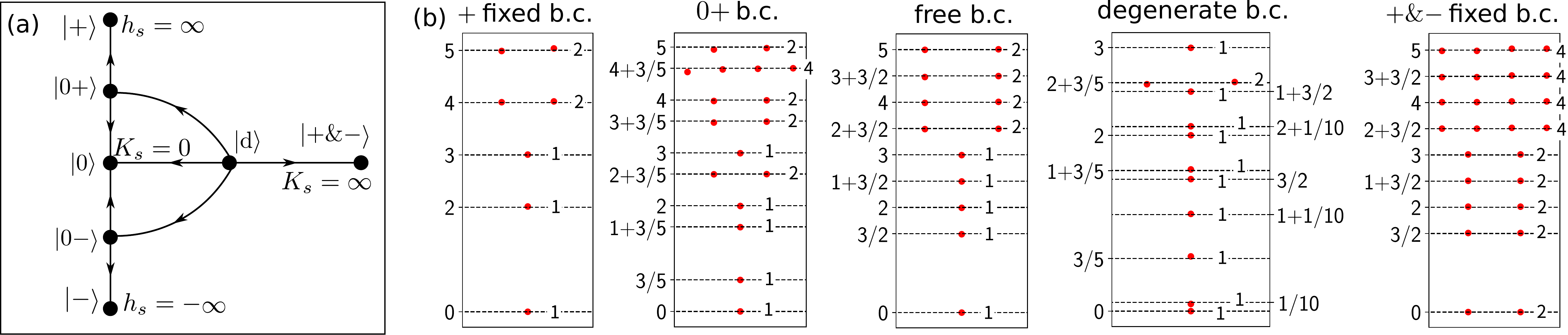}%
\caption{\label{fig:BC}
(Color online) (a) The boundary phase diagram of \eq{eq:BC-hamiltonian}. $\Ket{\mathrm{d}}$ represents the degenerated boundary states. (b) The computed boundary conformal spectrum by BTNR with $\chi=30$ extracted at the eighth RG step for the various boundary fixed points. The dashed line and the small figure near the plot again represent the exact value of the conformal dimensions and the exact degeneracy, which result from the operator content in Tab.~\ref{tab:TCI}.}
\end{figure*}
The physical meanings of each Cardy states are investigated by Chim, and the RG picture between them and the phase diagram are studied by Affleck~\cite{Chim1996,Affleck2000}. The remarkable point in the boundary phase diagram of this model in Fig.~\ref{fig:BC} (a) is the fixed point of the free b.c. and all of the $+$, $-$, and $+\&-$ fixed b.c. are stable, which results in the boundary phase transitions at a finite $h_s$ and a finite $K_s$, respectively. Notice that the free boundary state corresponds to $\Ket{0}$, where the boundary is dominated by the voids (i.e., $\sigma=0$ state). The scenario discussed analytically by BCFT is numerically investigated by Bl\"ote and Deng through the Monte Carlo simulation of \eq{eq:BC-hamiltonian}~\cite{Deng2004,Deng2005,Deng2019}. They detected the boundary phase transitions as conjectured by Affleck and confirmed that the critical exponents are quantitatively consistent with the conformal dimensions of the relevant fields in Tab.~\ref{tab:TCI}.

In this section, we show the conformal spectrum for each boundary fixed point by the tesor network method, which can be the complete numerical evidence of the Affleck's scenario in contrast to the Monte Carlo study which can investigate only the relevant scaling dimensions through the analyses of the order parameters. Though the conformal spectrum for the five of the six Cardy states are computed in the context of the quantum dimer model with a magnetic field~\cite{Chepiga2019}, the consistency between the BCFT conjecture and the spectrum in the lattice model for the degenerated boundary condition $\Ket{\mathrm{d}}$ is still not confirmed. In addition to the results for every Cardy's fixed point, we show the spectrum for the $+\&-$ fixed boundary condition.

We perform the numerical simulation in the following parameters as is also described in Tab.~\ref{tab:TCI}: the spectrum for the free b.c. is computed at $K_s=0$ and $h_s=0$, for the degenerate b.c. computed at $K_s=1.56623\times\Kcbulk$ and $h_s=0$, for the $0+$ b.c. computed at $K_s=\Kcbulk$ and $h_s=0.67721$, for the $+$ fixed b.c. computed at $K_s=\Kcbulk$ and $h_s=\infty$, and for the $+\&-$ fixed b.c. computed at $K_s=\infty$ and $h_s=0$. Those transition points are obtained by the Monte Carlo simulation~\cite{Deng2004}. We have not computed the spectrum for the $-$ fixed b.c. and $0-$ b.c., since the results are trivially the same as for the $+$ fixed b.c. and $0+$ b.c. due to the symmetry of the spin flip $\sigma\rightarrow -\sigma$. For the simulation of $h_s=0$, we can employ the $Z_2$ symmetric tensor as the simulation of the Ising model. The results are shown in Fig.~\ref{fig:BC} (b), all of which are completely consistent with the tri-ciritical Ising BCFT conjectures.

\subsection{3-state Potts model}

The Hamiltonian of the 3-state Potts model we utilize in this study is
\begin{eqnarray}
  \beta\mathcal{H}=-\Kcbulk\sum_{\langle ij\rangle\in\mathrm{bulk}}\delta_{\sigma_i\sigma_j}
  &-&K_s\sum_{\langle ij\rangle\in\mathrm{surface}}\delta_{\sigma_i\sigma_j}\nn
  &-&h_s\sum_{i\in\mathrm{surface}}\delta_{\sigma_iA},
  \label{eq:3Potts-hamiltonian}
\end{eqnarray}
where $\sigma=A$, $B$, or $C$ and $\Kcbulk=\ln\left(1+\sqrt{3}\right)$~\cite{Potts1952}.

Since the CFT of the Potts model is known to be the unitary minimal model $\mathcal{M}_{6,5}$ with the higher symmetry, the $\mathcal{W}_3$ symmetry, only the following six primary fields occur in the Potts BCFT which are composed of the subgroup of those in $\mathcal{M}_{6,5}$~\cite{Fateev1987}: $\phi_0$, $\phi_{\frac{2}{5}}$, $\phi_{\frac{7}{5}}$, $\phi_{3}$, $\phi_{\frac{1}{15}}$, and $\phi_{\frac{2}{3}}$, whose subscripts represent the conformal dimensions of the primary fields. Moreover, the two of the six primary fields with the non-zero $Z_3$ charge, $\phi_{\frac{1}{15}}$ and $\phi_{\frac{2}{3}}$, appear twice in the CFT spectrum, which have the opposite charges $+1$ and $-1$ respectively and are differentiated by being daggered like $\sigma$ and $\sigma^{\dagger}$. One can define the character of the extended algebra as
\begin{eqnarray}
  C_{\boldsymbol 1} &=& \chi_{0}+\chi_{3}\\
  C_{\epsilon} &=& \chi_{\frac{2}{5}}+\chi_{\frac{7}{5}}\\
  C_{\sigma} &=& C_{\sigma^{\dagger}} = \chi_{\frac{1}{15}}\\
  C_{\psi} &=& C_{\psi^{\dagger}} = \chi_{\frac{2}{3}},
\end{eqnarray}
where $C$ is the character for the $\mathcal{W}_3$ algebra and $\chi$ is the minimal character whose subscript means the conformal dimensions. Notice that the torus partition function of the 3-state Potts model becomes diagonal under the description of the $\mathcal{W}_3$ characters as
\begin{equation}
  Z = {C_{\boldsymbol 1}}^2+{C_{\epsilon}}^2+{C_{\sigma}}^2+{C_{\sigma^{\dagger}}}^2+{C_{\psi}}^2+{C_{\psi^{\dagger}}}^2.
\end{equation}

The list of the boundary fixed points and their operator contents are shown in Tab.~\ref{tab:3Potts}. While the first three states, $\Ket{A}$, $\Ket{B}$, and $\Ket{C}$, represent the fixed b.c.'s with a single spin, the next three states are also fixed b.c.'s with two types of spins: for instance, $\Ket{BC}$ means the disordered boundary state occupied by only $B$ and $C$. Though there are the six primary fields in the Potts BCFT as explained above, the whole set of the primary fields in $\mathcal{M}_{6,5}$ has to be taken into consideration to obtain the nontrivial Cardy states $\Ket{\mathrm{free}}$ and $\Ket{\mathrm{new}}$~\cite{Affleck1998,Fuchs1998}. The way to calculate the operator contents from the fusion rules is also explained in Ref.~\onlinecite{Affleck1998}.
\begin{table}[htb]
\caption{The Cardy states of the 3-state Potts model and the operator contents of the corresponding boundary fixed points. $(K_s,h_s)$ represents the location in the parameter space where the numerical computation has been performed.}
\label{tab:3Potts}
\begin{ruledtabular}
\begin{tabular}{ccc}
  Cardy state & operator content & $(K_s,h_s)$ \\\hline
  $\Ket{A}\equiv\cardy{\boldsymbol 1}$ & & $(0,+\infty)$ \\
  $\Ket{B}\equiv\cardy{\psi}$ & ${\boldsymbol 1}$ & \\
  $\Ket{C}\equiv\cardyw{\psi^{\dagger}}$ & & \\\hline
  $\Ket{BC}\equiv\cardy{\epsilon}$ & & $(0,-\infty)$ \\
  $\Ket{CA}\equiv\cardy{\sigma}$ & ${\boldsymbol 1}\oplus\epsilon$ & \\
  $\Ket{AB}\equiv\cardyw{\sigma^{\dagger}}$ & & \\\hline
  $\Ket{\mathrm{free}}$ & ${\boldsymbol 1}\oplus\psi\oplus\psi^{\dagger}$ & $(0,0)$ \\\hline
  $\Ket{\mathrm{new}}$ &
  \begin{tabular}{c}${\boldsymbol 1}\oplus\epsilon\oplus\sigma\oplus\sigma^{\dagger}$\\
    $\oplus\psi\oplus\psi^{\dagger}$\end{tabular}
  & \eq{eq:new-weight} \\\hline
      \begin{tabular}{l}$\Ket{A\& B\& C}$\\$\equiv\Ket{A}+\Ket{B}+\Ket{C}$\end{tabular}
      & $3\left({\boldsymbol 1}\oplus\psi\oplus\psi^{\dagger}\right)$ & $(+\infty,0)$\\
\end{tabular}
\end{ruledtabular}
\end{table}

\begin{figure*}
\includegraphics[width=18cm]{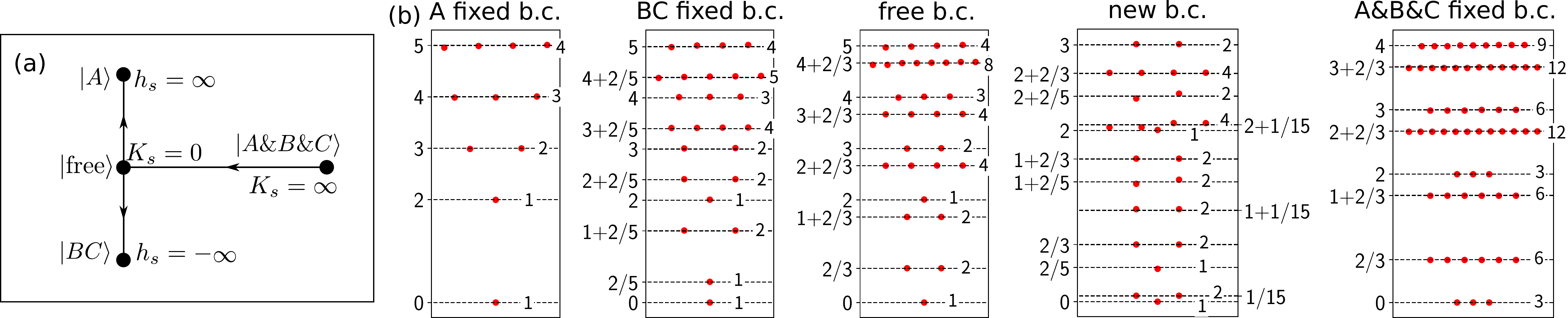}%
\caption{\label{fig:q3_phasediagram}
(Color online) (a) The boundary phase diagram of \eq{eq:3Potts-hamiltonian}. Note that the fixed point of the `new' b.c. is out of this phase diagram. (b) The computed boundary conformal spectrum by BTNR with $\chi=30$ at the eighth step for the various boundary fixed points.}
\end{figure*}

In Fig.~\ref{fig:q3_phasediagram} (a), we show the boundary phase diagram of the 3-state Potts model. Just like the Ising model, $K_s=0$ and $h_s=0$ yields the free b.c., which is stable for the direction of the positive $K_s$ but unstable for an infinitesimal magnetic field. The infinite surface coupling causes the spontaneously ordered boundary labeled as $\Ket{A\& B\& C}\equiv\Ket{A}+\Ket{B}+\Ket{C}$. The positive $h_s$ induces the fixed boundary state with a single spin $A$, while the negative $h_s$ causes the other fixed one occupied by two spins $B$ and $C$ with the same weight. The b.c. labeled as `new' does not appear in the phase diagram, since it can be realized only under some nonphysical condition~\cite{Affleck1998}. Using correspondence to the critical unitary A-D-E lattice models, Behrend and Pearce revealed~\cite{Behrend2001} that this boundary condition can be realized on the 3-state Potts model when the boundary Boltzmann weight matrix is
\begin{equation}
  \label{eq:new-weight}
  e^{K_s\delta_{\sigma_i\sigma_j}+\mathrm{Const.}}\equiv
  \begin{pmatrix}
    1 & -\frac{1}{2} & -\frac{1}{2} \\
    -\frac{1}{2} & 1 & -\frac{1}{2} \\
    -\frac{1}{2} & -\frac{1}{2} & 1 \\
  \end{pmatrix},
\end{equation}
which means that if the two neighbouring spins are the same the weight is $1$, while $-\frac{1}{2}$ if they are different. Although the negative Boltzmann weight causes the negative sign problem in Monte Carlo simulations, the tensor network methods work without any problem even in this case. We believe that the direct computation of the conformal spectrum with the new b.c. on the lattice is achieved first in our present work, although the Affleck-Ludwig boundary entropy with this b.c. is numerically obtained indirectly through the Kramers-Wannier duality~\cite{Tang2017}.

We show the computation results of the conformal spectrum of each boundary fixed point in Tab.~\ref{tab:3Potts}, in Fig.~\ref{fig:q3_phasediagram} (b). The computation is performed with the parameters right at the fixed points described in Fig.~\ref{fig:q3_phasediagram} (a) and \eq{eq:new-weight}, as described in Tab.~\ref{tab:3Potts}. We can adopt the $Z_3$ symmetric tensor for the simulation of $h_s=0$ and the $Z_2$ symmetric tensor even for $h_s\neq0$. We can confirm that all the results are consistent with the conjectures of the 3-state Potts BCFT.
%

\section{Conclusion\label{sec:conclusion}}

In this study, we compute the boundary conformal spectrum for the complete set of the boundary fixed points related to the Cardy states, for some simple classical spin systems, Ising, tri-critical Ising, and 3-state Potts model. As a numerical method capable of computing accurate spectrum, we implement the TNR algorithm for the system with open boundaries. We check that the BTNR algorithm, the TNR for the open-boundary system, generates the correct fixed point tensor even at criticality and the stable and accurate flow of the scaling dimensions. The numerical results obtained by BTNR are all consistent with the BCFT analysis and the surface critical behavior, which allows us to confirm the validity of the BCFT analyses for the lattice models.

The program presented in this paper can be extended to other cases where physical realization of the Cardy states in lattice models has not been fully established. For instance, in Ref.~\onlinecite{Deng2004,Deng2005} the authors investigated the surface critical behaviors of not only tri-critical Ising but also tri-critical 3-state Potts model, whose CFT corresponds to the next simplest minimal model $\mathcal{M}_{7,6}$ with an extended symmetry~\cite{Zamolodchikov1987}. The Cardy states of the tri-critical 3-state Potts model, however, are still not so investigated analytically. The numerical analysis used in this study will help us consider the operator contents of each boundary fixed point in the phase diagram obtained by the Monte Carlo simulation~\cite{Deng2004,Deng2005}.
%

\appendix*

\section{Boundary super operator\label{sec:sop}}

In this appendix, we present a method for extracting the conformal spectrum, diagonalization of the super operator, which is an alternative to the use of \eq{eq:spectrum-formula}. The similar method is useful to compute the bulk conformal spectrum by utilizing the ordinary TNR as shown in Ref.~\onlinecite{Evenbly2016}, which is based on the local conformal transformations implemented for lattices through the coarse graining procedure of TNR. This remarkable feature of TNR makes it possible to obtain not only the central charge and the conformal spectrum but also the coefficients of the operator product expansion between the operators in CFT.

\begin{figure}
\includegraphics[width=8.5cm]{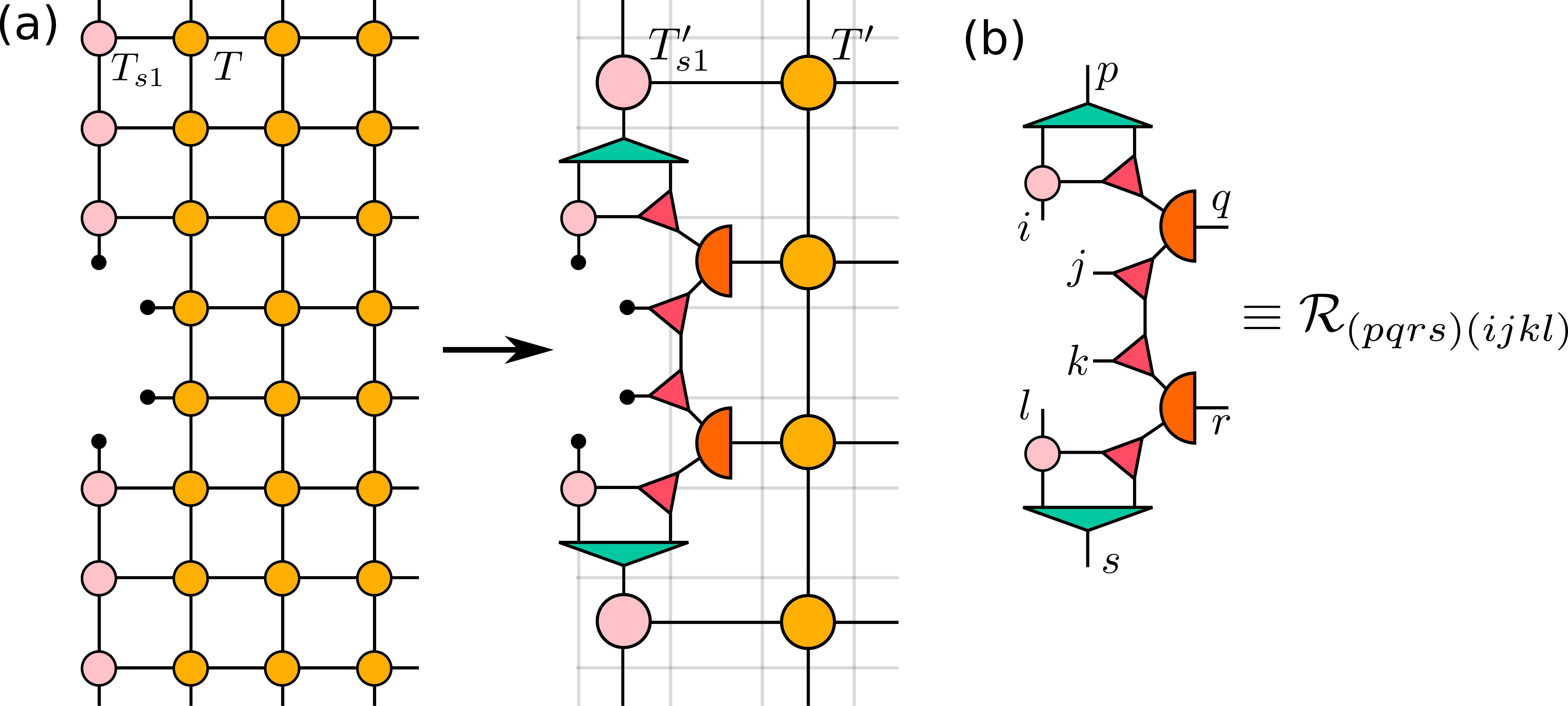}%
\caption{\label{fig:superop}
(Color online) (a) Perform the RG procedure of BTNR for the square lattice where the two neighbouring boundary tensors are removed. This yields the updated network constructed from $T'$ and $T_{s1}'$ with the leftovers made of the intermediates, which can be regarded as a super operator of the boundary tensors. (b) The super operator derived in (a), a rank-8 tensor, is considered as the matrix $\mathcal{R}$, the diagonalization of which amounts to the conformal spectrum.}
\end{figure}

Here, we demonstrate that one can construct super operators for boundary tensors by BTNR algorithm, which plays a roll of renormalizing the input tensors and doubling the length scale of them. The derivation of the boundary super operator is similar to that of the bulk version explained in Ref.~\onlinecite{Evenbly2016}. As shown in Fig.~\ref{fig:superop} (a), we consider a tensor network on the square lattice with the boundary, where the two adjacent boundary tensors are removed. Application of BTNR algorithm for this geometry produces some tensors around the hole in the boundary which are the remaining intermediates, in addition to the renormalized network with the twice length scale. These remaining tensors can be regarded as the super operator $\mathcal{R}$ as depicted in Fig.~\ref{fig:superop} (b), since this operator represents the RG map from the two boundary tensors $T_{s1}$ into the pair of $T'_{s1}$.

Because the super operator $\mathcal{R}$ implements the dilatation transformation on the lattice with boundaries, generated by the Virasoro operator $\hat{L}_0$, the eigen states of it are the primary fields and their descendants in CFT:
\begin{equation}
  \label{eq:superop}
  \mathcal{R}\Ket{\phi_k}=2^{-h_{\phi_k}}\Ket{\phi_k},
\end{equation}
where $\Ket{\phi_k}$ is the eigenvector of $\mathcal{R}$ corresponding to the operator $\phi_k$ in CFT, and $h_{\phi_k}$ is the conformal dimension of the operator $\phi_k$. Notice that the boundary tensors required for the construction of the super operator should be normalized so as to be scale-invariant~\cite{Iino2019}, and the super operator is also normalized so that the largest eigenvalue is $2^{-0}=1$ when the least conformal dimension is equal to zero, similarly to the bulk case~\cite{Evenbly2016}.

In Tab.~\ref{tab:spop_free}, we show the numerical results of the Ising model \eq{eq:potts-hamiltonian} with $K_s=h_s=0$, computed by \eq{eq:superop} with $\chi=16$ at the fifth RG step. Comparing with the exact values of the scaling dimensions, we can confirm that the obtained results are consistent with the operator content of the Ising BCFT with the free b.c.'s. The drawback of this method is, however, the construction and diagonalization of the super operator $\mathcal{R}$ requires highly expensive computational resources. Therefore, in the main text, we adopt the diagonalization of the transfer matrix \eq{eq:BCFTZ}, where larger bond dimension is available compared to the case of the super operator.
\begin{table}[htb]
\caption{The spectrum of the 8 smallest scaling dimensions except for the lowest one (i.e., 0) by the diagonalization of the super operator $\mathcal{R}$ with $\chi=16$ at the fifth RG step.}
\label{tab:spop_free}
\begin{ruledtabular}
  \begin{tabular}{ccccccccc}
    exact & 0.5 & 1.5 & 2 & 2.5 & 3 & 3.5 & 4 & 4 \\
    BTNR & 0.474 & 1.518 & 1.983 & 2.482 & 2.923 & 3.313 & 3.859 & 3.998\\
  \end{tabular}
\end{ruledtabular}
\end{table}
%


\begin{acknowledgments}
S.I. thanks Tateki Obori and Yoshiki Fukusumi for fruitful discussions, and is also grateful to the support of Program for Leading Graduate Schools (ALPS). N.K.'s work is financially supported by MEXT Grant-in-Aid for Scientific Research (B) (25287097, 19H01809). This research was supported by MEXT as "Exploratory Challenge on Post-K computer" (Frontiers of Basic Science: Challenging the Limits).
\end{acknowledgments}

\bibliography{bibliography}

\providecommand{\noopsort}[1]{}\providecommand{\singleletter}[1]{#1}%
\begin{thebibliography}{53}%
\makeatletter
\providecommand \@ifxundefined [1]{%
 \@ifx{#1\undefined}
}%
\providecommand \@ifnum [1]{%
 \ifnum #1\expandafter \@firstoftwo
 \else \expandafter \@secondoftwo
 \fi
}%
\providecommand \@ifx [1]{%
 \ifx #1\expandafter \@firstoftwo
 \else \expandafter \@secondoftwo
 \fi
}%
\providecommand \natexlab [1]{#1}%
\providecommand \enquote  [1]{``#1''}%
\providecommand \bibnamefont  [1]{#1}%
\providecommand \bibfnamefont [1]{#1}%
\providecommand \citenamefont [1]{#1}%
\providecommand \href@noop [0]{\@secondoftwo}%
\providecommand \href [0]{\begingroup \@sanitize@url \@href}%
\providecommand \@href[1]{\@@startlink{#1}\@@href}%
\providecommand \@@href[1]{\endgroup#1\@@endlink}%
\providecommand \@sanitize@url [0]{\catcode `\\12\catcode `\$12\catcode
  `\&12\catcode `\#12\catcode `\^12\catcode `\_12\catcode `\%12\relax}%
\providecommand \@@startlink[1]{}%
\providecommand \@@endlink[0]{}%
\providecommand \url  [0]{\begingroup\@sanitize@url \@url }%
\providecommand \@url [1]{\endgroup\@href {#1}{\urlprefix }}%
\providecommand \urlprefix  [0]{URL }%
\providecommand \Eprint [0]{\href }%
\providecommand \doibase [0]{http://dx.doi.org/}%
\providecommand \selectlanguage [0]{\@gobble}%
\providecommand \bibinfo  [0]{\@secondoftwo}%
\providecommand \bibfield  [0]{\@secondoftwo}%
\providecommand \translation [1]{[#1]}%
\providecommand \BibitemOpen [0]{}%
\providecommand \bibitemStop [0]{}%
\providecommand \bibitemNoStop [0]{.\EOS\space}%
\providecommand \EOS [0]{\spacefactor3000\relax}%
\providecommand \BibitemShut  [1]{\csname bibitem#1\endcsname}%
\let\auto@bib@innerbib\@empty
\bibitem [{\citenamefont {Belavin}\ \emph {et~al.}(1984)\citenamefont
  {Belavin}, \citenamefont {Polyakov},\ and\ \citenamefont
  {Zamolodchikov}}]{Belavin1984}%
  \BibitemOpen
  \bibfield  {author} {\bibinfo {author} {\bibfnamefont {A.~A.}\ \bibnamefont
  {Belavin}}, \bibinfo {author} {\bibfnamefont {A.~M.}\ \bibnamefont
  {Polyakov}}, \ and\ \bibinfo {author} {\bibfnamefont {A.~B.}\ \bibnamefont
  {Zamolodchikov}},\ }\href {\doibase 10.1016/0550-3213(84)90052-X} {\bibfield
  {journal} {\bibinfo  {journal} {Nuclear Physics B}\ }\textbf {\bibinfo
  {volume} {241}},\ \bibinfo {pages} {333} (\bibinfo {year}
  {1984})}\BibitemShut {NoStop}%
\bibitem [{\citenamefont {Francesco}\ \emph {et~al.}(1997)\citenamefont
  {Francesco}, \citenamefont {Mathieu},\ and\ \citenamefont
  {S{\'e}n{\'e}chal}}]{Francesco_CFT}%
  \BibitemOpen
  \bibfield  {author} {\bibinfo {author} {\bibfnamefont {P.~D.}\ \bibnamefont
  {Francesco}}, \bibinfo {author} {\bibfnamefont {P.}~\bibnamefont {Mathieu}},
  \ and\ \bibinfo {author} {\bibfnamefont {D.}~\bibnamefont
  {S{\'e}n{\'e}chal}},\ }\href@noop {} {\emph {\bibinfo {title} {Conformal
  Field Theory}}}\ (\bibinfo  {publisher} {Springer, New York},\ \bibinfo
  {year} {1997})\BibitemShut {NoStop}%
\bibitem [{\citenamefont {Cardy}(1984)}]{Cardy1984}%
  \BibitemOpen
  \bibfield  {author} {\bibinfo {author} {\bibfnamefont {J.~L.}\ \bibnamefont
  {Cardy}},\ }\href {\doibase 10.1016/0550-3213(84)90241-4} {\bibfield
  {journal} {\bibinfo  {journal} {Nuclear Physics B}\ }\textbf {\bibinfo
  {volume} {240}},\ \bibinfo {pages} {514} (\bibinfo {year}
  {1984})}\BibitemShut {NoStop}%
\bibitem [{\citenamefont {Cardy}(1989)}]{CARDY1989581}%
  \BibitemOpen
  \bibfield  {author} {\bibinfo {author} {\bibfnamefont {J.~L.}\ \bibnamefont
  {Cardy}},\ }\href {\doibase https://doi.org/10.1016/0550-3213(89)90521-X}
  {\bibfield  {journal} {\bibinfo  {journal} {Nuclear Physics B}\ }\textbf
  {\bibinfo {volume} {324}},\ \bibinfo {pages} {581 } (\bibinfo {year}
  {1989})}\BibitemShut {NoStop}%
\bibitem [{\citenamefont {Cardy}(1986)}]{CARDY1986200}%
  \BibitemOpen
  \bibfield  {author} {\bibinfo {author} {\bibfnamefont {J.~L.}\ \bibnamefont
  {Cardy}},\ }\href {\doibase https://doi.org/10.1016/0550-3213(86)90596-1}
  {\bibfield  {journal} {\bibinfo  {journal} {Nuclear Physics B}\ }\textbf
  {\bibinfo {volume} {275}},\ \bibinfo {pages} {200 } (\bibinfo {year}
  {1986})}\BibitemShut {NoStop}%
\bibitem [{\citenamefont {Binder}(1983)}]{Binder1983_DG}%
  \BibitemOpen
  \bibfield  {author} {\bibinfo {author} {\bibfnamefont {K.}~\bibnamefont
  {Binder}},\ }in\ \href@noop {} {\emph {\bibinfo {booktitle} {Phase
  Transitions and Critical Phenomena}}},\ Vol.~\bibinfo {volume} {8},\ \bibinfo
  {editor} {edited by\ \bibinfo {editor} {\bibfnamefont {C.}~\bibnamefont
  {Domb}}\ and\ \bibinfo {editor} {\bibfnamefont {J.~L.}\ \bibnamefont
  {Lebowitz}}}\ (\bibinfo  {publisher} {Academic Press, London},\ \bibinfo
  {year} {1983})\ p.~\bibinfo {pages} {1}\BibitemShut {NoStop}%
\bibitem [{\citenamefont {McCoy}\ and\ \citenamefont {Wu}(1967)}]{McCoy1967}%
  \BibitemOpen
  \bibfield  {author} {\bibinfo {author} {\bibfnamefont {B.~M.}\ \bibnamefont
  {McCoy}}\ and\ \bibinfo {author} {\bibfnamefont {T.~T.}\ \bibnamefont {Wu}},\
  }\href {\doibase 10.1103/PhysRev.162.436} {\bibfield  {journal} {\bibinfo
  {journal} {Phys. Rev.}\ }\textbf {\bibinfo {volume} {162}},\ \bibinfo {pages}
  {436} (\bibinfo {year} {1967})}\BibitemShut {NoStop}%
\bibitem [{\citenamefont {Yang}(1952)}]{Yang1952}%
  \BibitemOpen
  \bibfield  {author} {\bibinfo {author} {\bibfnamefont {C.~N.}\ \bibnamefont
  {Yang}},\ }\href {\doibase 10.1103/PhysRev.85.808} {\bibfield  {journal}
  {\bibinfo  {journal} {Phys. Rev.}\ }\textbf {\bibinfo {volume} {85}},\
  \bibinfo {pages} {808} (\bibinfo {year} {1952})}\BibitemShut {NoStop}%
\bibitem [{\citenamefont {Chim}(1996)}]{Chim1996}%
  \BibitemOpen
  \bibfield  {author} {\bibinfo {author} {\bibfnamefont {L.}~\bibnamefont
  {Chim}},\ }\href {\doibase 10.1142/S0217751X9600208X} {\bibfield  {journal}
  {\bibinfo  {journal} {International Journal of Modern Physics A}\ }\textbf
  {\bibinfo {volume} {11}},\ \bibinfo {pages} {4491} (\bibinfo {year}
  {1996})},\ \Eprint {http://arxiv.org/abs/hep-th/9510008}
  {arXiv:hep-th/9510008} \BibitemShut {NoStop}%
\bibitem [{\citenamefont {Affleck}\ \emph {et~al.}(1998)\citenamefont
  {Affleck}, \citenamefont {Oshikawa},\ and\ \citenamefont
  {Saleur}}]{Affleck1998}%
  \BibitemOpen
  \bibfield  {author} {\bibinfo {author} {\bibfnamefont {I.}~\bibnamefont
  {Affleck}}, \bibinfo {author} {\bibfnamefont {M.}~\bibnamefont {Oshikawa}}, \
  and\ \bibinfo {author} {\bibfnamefont {H.}~\bibnamefont {Saleur}},\ }\href
  {\doibase 10.1088/0305-4470/31/28/003} {\bibfield  {journal} {\bibinfo
  {journal} {Journal of Physics A: Mathematical and General}\ }\textbf
  {\bibinfo {volume} {31}},\ \bibinfo {pages} {5827} (\bibinfo {year}
  {1998})},\ \Eprint {http://arxiv.org/abs/cond-mat/9804117}
  {arXiv:cond-mat/9804117} \BibitemShut {NoStop}%
\bibitem [{\citenamefont {Behrend}\ and\ \citenamefont
  {Pearce}(2001)}]{Behrend2001}%
  \BibitemOpen
  \bibfield  {author} {\bibinfo {author} {\bibfnamefont {R.~E.}\ \bibnamefont
  {Behrend}}\ and\ \bibinfo {author} {\bibfnamefont {P.~A.}\ \bibnamefont
  {Pearce}},\ }\href {\doibase 10.1023/A:1004890600991} {\bibfield  {journal}
  {\bibinfo  {journal} {Journal of Statistical Physics}\ }\textbf {\bibinfo
  {volume} {102}},\ \bibinfo {pages} {577} (\bibinfo {year} {2001})},\ \Eprint
  {http://arxiv.org/abs/hep-th/0006094} {arXiv:hep-th/0006094} \BibitemShut
  {NoStop}%
\bibitem [{\citenamefont {Burkhardt}\ and\ \citenamefont
  {Guim}(1985)}]{Burkhardt1985}%
  \BibitemOpen
  \bibfield  {author} {\bibinfo {author} {\bibfnamefont {T.~W.}\ \bibnamefont
  {Burkhardt}}\ and\ \bibinfo {author} {\bibfnamefont {I.}~\bibnamefont
  {Guim}},\ }\href {\doibase 10.1088/0305-4470/18/1/006} {\bibfield  {journal}
  {\bibinfo  {journal} {Journal of Physics A: Mathematical and General}\
  }\textbf {\bibinfo {volume} {18}},\ \bibinfo {pages} {33} (\bibinfo {year}
  {1985})}\BibitemShut {NoStop}%
\bibitem [{\citenamefont {Evenbly}\ \emph {et~al.}(2010)\citenamefont
  {Evenbly}, \citenamefont {Pfeifer}, \citenamefont {Pic\'o}, \citenamefont
  {Iblisdir}, \citenamefont {Tagliacozzo}, \citenamefont {McCulloch},\ and\
  \citenamefont {Vidal}}]{Evenbly2010}%
  \BibitemOpen
  \bibfield  {author} {\bibinfo {author} {\bibfnamefont {G.}~\bibnamefont
  {Evenbly}}, \bibinfo {author} {\bibfnamefont {R.~N.~C.}\ \bibnamefont
  {Pfeifer}}, \bibinfo {author} {\bibfnamefont {V.}~\bibnamefont {Pic\'o}},
  \bibinfo {author} {\bibfnamefont {S.}~\bibnamefont {Iblisdir}}, \bibinfo
  {author} {\bibfnamefont {L.}~\bibnamefont {Tagliacozzo}}, \bibinfo {author}
  {\bibfnamefont {I.~P.}\ \bibnamefont {McCulloch}}, \ and\ \bibinfo {author}
  {\bibfnamefont {G.}~\bibnamefont {Vidal}},\ }\href {\doibase
  10.1103/PhysRevB.82.161107} {\bibfield  {journal} {\bibinfo  {journal} {Phys.
  Rev. B}\ }\textbf {\bibinfo {volume} {82}},\ \bibinfo {pages} {161107(R)}
  (\bibinfo {year} {2010})},\ \Eprint {http://arxiv.org/abs/0912.1642}
  {arXiv:0912.1642 [cond-mat.str-el]} \BibitemShut {NoStop}%
\bibitem [{\citenamefont {Balaska}\ and\ \citenamefont
  {Bounoua}(2013)}]{Balaska2013}%
  \BibitemOpen
  \bibfield  {author} {\bibinfo {author} {\bibfnamefont {S.}~\bibnamefont
  {Balaska}}\ and\ \bibinfo {author} {\bibfnamefont {N.~S.}\ \bibnamefont
  {Bounoua}},\ }\href {\doibase 10.1088/1742-6596/411/1/012004} {\bibfield
  {journal} {\bibinfo  {journal} {Journal of Physics: Conference Series}\
  }\textbf {\bibinfo {volume} {411}},\ \bibinfo {pages} {012004} (\bibinfo
  {year} {2013})},\ \Eprint {http://arxiv.org/abs/1104.1104} {arXiv:1104.1104
  [cond-mat.stat-mech]} \BibitemShut {NoStop}%
\bibitem [{\citenamefont {von Gehlen}\ \emph {et~al.}(1986)\citenamefont {von
  Gehlen}, \citenamefont {Rittenberg},\ and\ \citenamefont
  {Ruegg}}]{Gehlen1986_2}%
  \BibitemOpen
  \bibfield  {author} {\bibinfo {author} {\bibfnamefont {G.}~\bibnamefont {von
  Gehlen}}, \bibinfo {author} {\bibfnamefont {V.}~\bibnamefont {Rittenberg}}, \
  and\ \bibinfo {author} {\bibfnamefont {H.}~\bibnamefont {Ruegg}},\ }\href
  {\doibase 10.1088/0305-4470/19/1/014} {\bibfield  {journal} {\bibinfo
  {journal} {Journal of Physics A: Mathematical and General}\ }\textbf
  {\bibinfo {volume} {19}},\ \bibinfo {pages} {107} (\bibinfo {year}
  {1986})}\BibitemShut {NoStop}%
\bibitem [{\citenamefont {von Gehlen}\ and\ \citenamefont
  {Rittenberg}(1986)}]{Gehlen1986}%
  \BibitemOpen
  \bibfield  {author} {\bibinfo {author} {\bibfnamefont {G.}~\bibnamefont {von
  Gehlen}}\ and\ \bibinfo {author} {\bibfnamefont {V.}~\bibnamefont
  {Rittenberg}},\ }\href {\doibase 10.1088/0305-4470/19/10/014} {\bibfield
  {journal} {\bibinfo  {journal} {Journal of Physics A: Mathematical and
  General}\ }\textbf {\bibinfo {volume} {19}},\ \bibinfo {pages} {L631}
  (\bibinfo {year} {1986})}\BibitemShut {NoStop}%
\bibitem [{\citenamefont {Balbao}\ and\ \citenamefont
  {de~Felicio}(1987)}]{Balbao1987}%
  \BibitemOpen
  \bibfield  {author} {\bibinfo {author} {\bibfnamefont {D.~B.}\ \bibnamefont
  {Balbao}}\ and\ \bibinfo {author} {\bibfnamefont {J.~R.~D.}\ \bibnamefont
  {de~Felicio}},\ }\href {\doibase 10.1088/0305-4470/20/4/005} {\bibfield
  {journal} {\bibinfo  {journal} {Journal of Physics A: Mathematical and
  General}\ }\textbf {\bibinfo {volume} {20}},\ \bibinfo {pages} {L207}
  (\bibinfo {year} {1987})}\BibitemShut {NoStop}%
\bibitem [{\citenamefont {L{\"a}uchli}(2013)}]{Lauchli2013}%
  \BibitemOpen
  \bibfield  {author} {\bibinfo {author} {\bibfnamefont {A.~M.}\ \bibnamefont
  {L{\"a}uchli}},\ }\href@noop {} {\  (\bibinfo {year} {2013})},\ \Eprint
  {http://arxiv.org/abs/1303.0741} {arXiv:1303.0741 [cond-mat.stat-mech]}
  \BibitemShut {NoStop}%
\bibitem [{\citenamefont {Chepiga}\ and\ \citenamefont
  {Mila}(2017)}]{Chepiga2017}%
  \BibitemOpen
  \bibfield  {author} {\bibinfo {author} {\bibfnamefont {N.}~\bibnamefont
  {Chepiga}}\ and\ \bibinfo {author} {\bibfnamefont {F.}~\bibnamefont {Mila}},\
  }\href {\doibase 10.1103/PhysRevB.96.054425} {\bibfield  {journal} {\bibinfo
  {journal} {Phys. Rev. B}\ }\textbf {\bibinfo {volume} {96}},\ \bibinfo
  {pages} {054425} (\bibinfo {year} {2017})},\ \Eprint
  {http://arxiv.org/abs/1705.05423} {arXiv:1705.05423 [cond-mat.str-el]}
  \BibitemShut {NoStop}%
\bibitem [{\citenamefont {Chepiga}\ and\ \citenamefont
  {Mila}(2019)}]{Chepiga2019}%
  \BibitemOpen
  \bibfield  {author} {\bibinfo {author} {\bibfnamefont {N.}~\bibnamefont
  {Chepiga}}\ and\ \bibinfo {author} {\bibfnamefont {F.}~\bibnamefont {Mila}},\
  }\href {\doibase 10.21468/SciPostPhys.6.3.033} {\bibfield  {journal}
  {\bibinfo  {journal} {SciPost Phys.}\ }\textbf {\bibinfo {volume} {6}},\
  \bibinfo {pages} {33} (\bibinfo {year} {2019})},\ \Eprint
  {http://arxiv.org/abs/1809.00746} {arXiv:1809.00746 [cond-mat.str-el]}
  \BibitemShut {NoStop}%
\bibitem [{\citenamefont {Evenbly}\ and\ \citenamefont
  {Vidal}(2015{\natexlab{a}})}]{Evenbly2015}%
  \BibitemOpen
  \bibfield  {author} {\bibinfo {author} {\bibfnamefont {G.}~\bibnamefont
  {Evenbly}}\ and\ \bibinfo {author} {\bibfnamefont {G.}~\bibnamefont
  {Vidal}},\ }\href {\doibase 10.1103/PhysRevLett.115.180405} {\bibfield
  {journal} {\bibinfo  {journal} {Phys. Rev. Lett.}\ }\textbf {\bibinfo
  {volume} {115}},\ \bibinfo {pages} {180405} (\bibinfo {year}
  {2015}{\natexlab{a}})},\ \Eprint {http://arxiv.org/abs/1412.0732}
  {arXiv:1412.0732 [cond-mat.str-el]} \BibitemShut {NoStop}%
\bibitem [{\citenamefont {Evenbly}(2017)}]{Evenbly2017}%
  \BibitemOpen
  \bibfield  {author} {\bibinfo {author} {\bibfnamefont {G.}~\bibnamefont
  {Evenbly}},\ }\href {\doibase 10.1103/PhysRevB.95.045117} {\bibfield
  {journal} {\bibinfo  {journal} {Phys. Rev. B}\ }\textbf {\bibinfo {volume}
  {95}},\ \bibinfo {pages} {45117} (\bibinfo {year} {2017})},\ \Eprint
  {http://arxiv.org/abs/1509.07484} {arXiv:1509.07484} \BibitemShut {NoStop}%
\bibitem [{\citenamefont {Iino}\ \emph {et~al.}(2019)\citenamefont {Iino},
  \citenamefont {Morita},\ and\ \citenamefont {Kawashima}}]{Iino2019}%
  \BibitemOpen
  \bibfield  {author} {\bibinfo {author} {\bibfnamefont {S.}~\bibnamefont
  {Iino}}, \bibinfo {author} {\bibfnamefont {S.}~\bibnamefont {Morita}}, \ and\
  \bibinfo {author} {\bibfnamefont {N.}~\bibnamefont {Kawashima}},\ }\href
  {\doibase 10.1103/PhysRevB.100.035449} {\bibfield  {journal} {\bibinfo
  {journal} {Phys. Rev. B}\ }\textbf {\bibinfo {volume} {100}},\ \bibinfo
  {pages} {35449} (\bibinfo {year} {2019})},\ \Eprint
  {http://arxiv.org/abs/1905.02351} {arXiv:1905.02351 [cond-mat.stat-mech]}
  \BibitemShut {NoStop}%
\bibitem [{\citenamefont {Levin}\ and\ \citenamefont {Nave}(2007)}]{Levin2007}%
  \BibitemOpen
  \bibfield  {author} {\bibinfo {author} {\bibfnamefont {M.}~\bibnamefont
  {Levin}}\ and\ \bibinfo {author} {\bibfnamefont {C.~P.}\ \bibnamefont
  {Nave}},\ }\href {\doibase 10.1103/PhysRevLett.99.120601} {\bibfield
  {journal} {\bibinfo  {journal} {Phys. Rev. Lett.}\ }\textbf {\bibinfo
  {volume} {99}},\ \bibinfo {pages} {120601} (\bibinfo {year}
  {2007})}\BibitemShut {NoStop}%
\bibitem [{\citenamefont {Gu}\ and\ \citenamefont {Wen}(2009)}]{Gu2009}%
  \BibitemOpen
  \bibfield  {author} {\bibinfo {author} {\bibfnamefont {Z.-C.}\ \bibnamefont
  {Gu}}\ and\ \bibinfo {author} {\bibfnamefont {X.-G.}\ \bibnamefont {Wen}},\
  }\href {\doibase 10.1103/PhysRevB.80.155131} {\bibfield  {journal} {\bibinfo
  {journal} {Phys. Rev. B}\ }\textbf {\bibinfo {volume} {80}},\ \bibinfo
  {pages} {155131} (\bibinfo {year} {2009})},\ \Eprint
  {http://arxiv.org/abs/0903.1069} {arXiv:0903.1069 [cond-mat.str-el]}
  \BibitemShut {NoStop}%
\bibitem [{\citenamefont {Xie}\ \emph {et~al.}(2009)\citenamefont {Xie},
  \citenamefont {Jiang}, \citenamefont {Chen}, \citenamefont {Weng},\ and\
  \citenamefont {Xiang}}]{Xie2009}%
  \BibitemOpen
  \bibfield  {author} {\bibinfo {author} {\bibfnamefont {Z.~Y.}\ \bibnamefont
  {Xie}}, \bibinfo {author} {\bibfnamefont {H.~C.}\ \bibnamefont {Jiang}},
  \bibinfo {author} {\bibfnamefont {Q.~N.}\ \bibnamefont {Chen}}, \bibinfo
  {author} {\bibfnamefont {Z.~Y.}\ \bibnamefont {Weng}}, \ and\ \bibinfo
  {author} {\bibfnamefont {T.}~\bibnamefont {Xiang}},\ }\href {\doibase
  10.1103/PhysRevLett.103.160601} {\bibfield  {journal} {\bibinfo  {journal}
  {Phys. Rev. Lett.}\ }\textbf {\bibinfo {volume} {103}},\ \bibinfo {pages}
  {160601} (\bibinfo {year} {2009})},\ \Eprint {http://arxiv.org/abs/0809.0182}
  {arXiv:0809.0182 [cond-mat.str-el]} \BibitemShut {NoStop}%
\bibitem [{\citenamefont {Zhao}\ \emph {et~al.}(2010)\citenamefont {Zhao},
  \citenamefont {Xie}, \citenamefont {Chen}, \citenamefont {Wei}, \citenamefont
  {Cai},\ and\ \citenamefont {Xiang}}]{Zhao2010}%
  \BibitemOpen
  \bibfield  {author} {\bibinfo {author} {\bibfnamefont {H.~H.}\ \bibnamefont
  {Zhao}}, \bibinfo {author} {\bibfnamefont {Z.~Y.}\ \bibnamefont {Xie}},
  \bibinfo {author} {\bibfnamefont {Q.~N.}\ \bibnamefont {Chen}}, \bibinfo
  {author} {\bibfnamefont {Z.~C.}\ \bibnamefont {Wei}}, \bibinfo {author}
  {\bibfnamefont {J.~W.}\ \bibnamefont {Cai}}, \ and\ \bibinfo {author}
  {\bibfnamefont {T.}~\bibnamefont {Xiang}},\ }\href {\doibase
  10.1103/PhysRevB.81.174411} {\bibfield  {journal} {\bibinfo  {journal} {Phys.
  Rev. B}\ }\textbf {\bibinfo {volume} {81}},\ \bibinfo {pages} {174411}
  (\bibinfo {year} {2010})},\ \Eprint {http://arxiv.org/abs/1002.1405}
  {arXiv:1002.1405 [cond-mat.str-el]} \BibitemShut {NoStop}%
\bibitem [{\citenamefont {Xie}\ \emph {et~al.}(2012)\citenamefont {Xie},
  \citenamefont {Chen}, \citenamefont {Qin}, \citenamefont {Zhu}, \citenamefont
  {Yang},\ and\ \citenamefont {Xiang}}]{Xie2012}%
  \BibitemOpen
  \bibfield  {author} {\bibinfo {author} {\bibfnamefont {Z.~Y.}\ \bibnamefont
  {Xie}}, \bibinfo {author} {\bibfnamefont {J.}~\bibnamefont {Chen}}, \bibinfo
  {author} {\bibfnamefont {M.~P.}\ \bibnamefont {Qin}}, \bibinfo {author}
  {\bibfnamefont {J.~W.}\ \bibnamefont {Zhu}}, \bibinfo {author} {\bibfnamefont
  {L.~P.}\ \bibnamefont {Yang}}, \ and\ \bibinfo {author} {\bibfnamefont
  {T.}~\bibnamefont {Xiang}},\ }\href {\doibase 10.1103/PhysRevB.86.045139}
  {\bibfield  {journal} {\bibinfo  {journal} {Phys. Rev. B}\ }\textbf {\bibinfo
  {volume} {86}},\ \bibinfo {pages} {045139} (\bibinfo {year} {2012})},\
  \Eprint {http://arxiv.org/abs/1201.1144} {arXiv:1201.1144
  [cond-mat.stat-mech]} \BibitemShut {NoStop}%
\bibitem [{\citenamefont {Yang}\ \emph {et~al.}(2017)\citenamefont {Yang},
  \citenamefont {Gu},\ and\ \citenamefont {Wen}}]{Yang2017}%
  \BibitemOpen
  \bibfield  {author} {\bibinfo {author} {\bibfnamefont {S.}~\bibnamefont
  {Yang}}, \bibinfo {author} {\bibfnamefont {Z.-C.}\ \bibnamefont {Gu}}, \ and\
  \bibinfo {author} {\bibfnamefont {X.-G.}\ \bibnamefont {Wen}},\ }\href
  {\doibase 10.1103/PhysRevLett.118.110504} {\bibfield  {journal} {\bibinfo
  {journal} {Phys. Rev. Lett.}\ }\textbf {\bibinfo {volume} {118}},\ \bibinfo
  {pages} {110504} (\bibinfo {year} {2017})},\ \Eprint
  {http://arxiv.org/abs/1512.04938} {arXiv:1512.04938 [cond-mat.str-el]}
  \BibitemShut {NoStop}%
\bibitem [{\citenamefont {Bal}\ \emph {et~al.}(2017)\citenamefont {Bal},
  \citenamefont {Mari\"en}, \citenamefont {Haegeman},\ and\ \citenamefont
  {Verstraete}}]{Bal2017}%
  \BibitemOpen
  \bibfield  {author} {\bibinfo {author} {\bibfnamefont {M.}~\bibnamefont
  {Bal}}, \bibinfo {author} {\bibfnamefont {M.}~\bibnamefont {Mari\"en}},
  \bibinfo {author} {\bibfnamefont {J.}~\bibnamefont {Haegeman}}, \ and\
  \bibinfo {author} {\bibfnamefont {F.}~\bibnamefont {Verstraete}},\ }\href
  {\doibase 10.1103/PhysRevLett.118.250602} {\bibfield  {journal} {\bibinfo
  {journal} {Phys. Rev. Lett.}\ }\textbf {\bibinfo {volume} {118}},\ \bibinfo
  {pages} {250602} (\bibinfo {year} {2017})},\ \Eprint
  {http://arxiv.org/abs/1703.00365} {arXiv:1703.00365 [cond-mat.stat-mech]}
  \BibitemShut {NoStop}%
\bibitem [{\citenamefont {Hauru}\ \emph {et~al.}(2018)\citenamefont {Hauru},
  \citenamefont {Delcamp},\ and\ \citenamefont {Mizera}}]{Hauru2018}%
  \BibitemOpen
  \bibfield  {author} {\bibinfo {author} {\bibfnamefont {M.}~\bibnamefont
  {Hauru}}, \bibinfo {author} {\bibfnamefont {C.}~\bibnamefont {Delcamp}}, \
  and\ \bibinfo {author} {\bibfnamefont {S.}~\bibnamefont {Mizera}},\ }\href
  {\doibase 10.1103/PhysRevB.97.045111} {\bibfield  {journal} {\bibinfo
  {journal} {Phys. Rev. B}\ }\textbf {\bibinfo {volume} {97}},\ \bibinfo
  {pages} {045111} (\bibinfo {year} {2018})},\ \Eprint
  {http://arxiv.org/abs/1709.07460} {arXiv:1709.07460 [cond-mat.str-el]}
  \BibitemShut {NoStop}%
\bibitem [{\citenamefont {Morita}\ \emph {et~al.}(2018)\citenamefont {Morita},
  \citenamefont {Igarashi}, \citenamefont {Zhao},\ and\ \citenamefont
  {Kawashima}}]{Morita2017}%
  \BibitemOpen
  \bibfield  {author} {\bibinfo {author} {\bibfnamefont {S.}~\bibnamefont
  {Morita}}, \bibinfo {author} {\bibfnamefont {R.}~\bibnamefont {Igarashi}},
  \bibinfo {author} {\bibfnamefont {H.-H.}\ \bibnamefont {Zhao}}, \ and\
  \bibinfo {author} {\bibfnamefont {N.}~\bibnamefont {Kawashima}},\ }\href@noop
  {} {\bibfield  {journal} {\bibinfo  {journal} {Phys. Rev. E}\ }\textbf
  {\bibinfo {volume} {97}},\ \bibinfo {pages} {033310} (\bibinfo {year}
  {2018})},\ \Eprint {http://arxiv.org/abs/1712.01458} {arXiv:1712.01458
  [cond-mat.stat-mech]} \BibitemShut {NoStop}%
\bibitem [{\citenamefont {Nakamura}\ \emph {et~al.}(2019)\citenamefont
  {Nakamura}, \citenamefont {Oba},\ and\ \citenamefont
  {Takeda}}]{Nakamura2019}%
  \BibitemOpen
  \bibfield  {author} {\bibinfo {author} {\bibfnamefont {Y.}~\bibnamefont
  {Nakamura}}, \bibinfo {author} {\bibfnamefont {H.}~\bibnamefont {Oba}}, \
  and\ \bibinfo {author} {\bibfnamefont {S.}~\bibnamefont {Takeda}},\ }\href
  {\doibase 10.1103/PhysRevB.99.155101} {\bibfield  {journal} {\bibinfo
  {journal} {Phys. Rev. B}\ }\textbf {\bibinfo {volume} {99}},\ \bibinfo
  {pages} {155101} (\bibinfo {year} {2019})},\ \Eprint
  {http://arxiv.org/abs/1809.08030} {arXiv:1809.08030 [cond-mat.stat-mech]}
  \BibitemShut {NoStop}%
\bibitem [{\citenamefont {Adachi}\ \emph {et~al.}(2019)\citenamefont {Adachi},
  \citenamefont {Okubo},\ and\ \citenamefont {Todo}}]{Adachi2019}%
  \BibitemOpen
  \bibfield  {author} {\bibinfo {author} {\bibfnamefont {D.}~\bibnamefont
  {Adachi}}, \bibinfo {author} {\bibfnamefont {T.}~\bibnamefont {Okubo}}, \
  and\ \bibinfo {author} {\bibfnamefont {S.}~\bibnamefont {Todo}},\ }\href@noop
  {} {\  (\bibinfo {year} {2019})},\ \Eprint {http://arxiv.org/abs/1906.02007}
  {arXiv:1906.02007 [cond-mat.stat-mech]} \BibitemShut {NoStop}%
\bibitem [{\citenamefont {Kadoh}\ and\ \citenamefont
  {Nakayama}(2019)}]{Kadoh2019}%
  \BibitemOpen
  \bibfield  {author} {\bibinfo {author} {\bibfnamefont {D.}~\bibnamefont
  {Kadoh}}\ and\ \bibinfo {author} {\bibfnamefont {K.}~\bibnamefont
  {Nakayama}},\ }\href@noop {} {\  (\bibinfo {year} {2019})},\ \Eprint
  {http://arxiv.org/abs/1912.02414} {arXiv:1912.02414 [hep-lat]} \BibitemShut
  {NoStop}%
\bibitem [{\citenamefont {Cardy}(1996)}]{Cardy_statmech}%
  \BibitemOpen
  \bibfield  {author} {\bibinfo {author} {\bibfnamefont {J.~L.}\ \bibnamefont
  {Cardy}},\ }\href@noop {} {\emph {\bibinfo {title} {Scaling and
  Renormalization in Statistical Physics}}}\ (\bibinfo  {publisher} {Cambridge
  University Press},\ \bibinfo {year} {1996})\BibitemShut {NoStop}%
\bibitem [{\citenamefont {Hauru}\ \emph {et~al.}(2016)\citenamefont {Hauru},
  \citenamefont {Evenbly}, \citenamefont {Ho}, \citenamefont {Gaiotto},\ and\
  \citenamefont {Vidal}}]{Hauru2016}%
  \BibitemOpen
  \bibfield  {author} {\bibinfo {author} {\bibfnamefont {M.}~\bibnamefont
  {Hauru}}, \bibinfo {author} {\bibfnamefont {G.}~\bibnamefont {Evenbly}},
  \bibinfo {author} {\bibfnamefont {W.~W.}\ \bibnamefont {Ho}}, \bibinfo
  {author} {\bibfnamefont {D.}~\bibnamefont {Gaiotto}}, \ and\ \bibinfo
  {author} {\bibfnamefont {G.}~\bibnamefont {Vidal}},\ }\href {\doibase
  10.1103/PhysRevB.94.115125} {\bibfield  {journal} {\bibinfo  {journal} {Phys.
  Rev. B}\ }\textbf {\bibinfo {volume} {94}},\ \bibinfo {pages} {115125}
  (\bibinfo {year} {2016})},\ \Eprint {http://arxiv.org/abs/1512.03846}
  {arXiv:1512.03846 [cond-mat.str-el]} \BibitemShut {NoStop}%
\bibitem [{\citenamefont {Evenbly}\ and\ \citenamefont
  {Vidal}(2015{\natexlab{b}})}]{Evenbly2015_2}%
  \BibitemOpen
  \bibfield  {author} {\bibinfo {author} {\bibfnamefont {G.}~\bibnamefont
  {Evenbly}}\ and\ \bibinfo {author} {\bibfnamefont {G.}~\bibnamefont
  {Vidal}},\ }\href {\doibase 10.1103/PhysRevLett.115.200401} {\bibfield
  {journal} {\bibinfo  {journal} {Phys. Rev. Lett.}\ }\textbf {\bibinfo
  {volume} {115}},\ \bibinfo {pages} {200401} (\bibinfo {year}
  {2015}{\natexlab{b}})},\ \Eprint {http://arxiv.org/abs/1502.05385}
  {arXiv:1502.05385 [cond-mat.str-el]} \BibitemShut {NoStop}%
\bibitem [{\citenamefont {Ising}(1925)}]{Ising1925}%
  \BibitemOpen
  \bibfield  {author} {\bibinfo {author} {\bibfnamefont {E.}~\bibnamefont
  {Ising}},\ }\href {\doibase 10.1007/BF02980577} {\bibfield  {journal}
  {\bibinfo  {journal} {Zeitschrift f{\"u}r Physik}\ }\textbf {\bibinfo
  {volume} {31}},\ \bibinfo {pages} {253} (\bibinfo {year} {1925})}\BibitemShut
  {NoStop}%
\bibitem [{\citenamefont {Singh}\ \emph {et~al.}(2011)\citenamefont {Singh},
  \citenamefont {Pfeifer},\ and\ \citenamefont {Vidal}}]{Singh2011}%
  \BibitemOpen
  \bibfield  {author} {\bibinfo {author} {\bibfnamefont {S.}~\bibnamefont
  {Singh}}, \bibinfo {author} {\bibfnamefont {R.~N.~C.}\ \bibnamefont
  {Pfeifer}}, \ and\ \bibinfo {author} {\bibfnamefont {G.}~\bibnamefont
  {Vidal}},\ }\href {\doibase 10.1103/PhysRevB.83.115125} {\bibfield  {journal}
  {\bibinfo  {journal} {Phys. Rev. B}\ }\textbf {\bibinfo {volume} {83}},\
  \bibinfo {pages} {115125} (\bibinfo {year} {2011})},\ \Eprint
  {http://arxiv.org/abs/1008.4774} {arXiv:1008.4774 [cond-mat.str-el]}
  \BibitemShut {NoStop}%
\bibitem [{\citenamefont {Affleck}(2000)}]{Affleck2000}%
  \BibitemOpen
  \bibfield  {author} {\bibinfo {author} {\bibfnamefont {I.}~\bibnamefont
  {Affleck}},\ }\href {\doibase 10.1088/0305-4470/33/37/301} {\bibfield
  {journal} {\bibinfo  {journal} {Journal of Physics A: Mathematical and
  General}\ }\textbf {\bibinfo {volume} {33}},\ \bibinfo {pages} {6473}
  (\bibinfo {year} {2000})},\ \Eprint {http://arxiv.org/abs/cond-mat/0005286}
  {arXiv:cond-mat/0005286} \BibitemShut {NoStop}%
\bibitem [{\citenamefont {Blume}(1966)}]{Blume1966}%
  \BibitemOpen
  \bibfield  {author} {\bibinfo {author} {\bibfnamefont {M.}~\bibnamefont
  {Blume}},\ }\href {\doibase 10.1103/PhysRev.141.517} {\bibfield  {journal}
  {\bibinfo  {journal} {Phys. Rev.}\ }\textbf {\bibinfo {volume} {141}},\
  \bibinfo {pages} {517} (\bibinfo {year} {1966})}\BibitemShut {NoStop}%
\bibitem [{\citenamefont {Capel}(1966)}]{Capel1966}%
  \BibitemOpen
  \bibfield  {author} {\bibinfo {author} {\bibfnamefont {H.}~\bibnamefont
  {Capel}},\ }\href {\doibase https://doi.org/10.1016/0031-8914(66)90027-9}
  {\bibfield  {journal} {\bibinfo  {journal} {Physica}\ }\textbf {\bibinfo
  {volume} {32}},\ \bibinfo {pages} {966 } (\bibinfo {year}
  {1966})}\BibitemShut {NoStop}%
\bibitem [{\citenamefont {Deng}\ and\ \citenamefont
  {Bl\"ote}(2004)}]{Deng2004}%
  \BibitemOpen
  \bibfield  {author} {\bibinfo {author} {\bibfnamefont {Y.}~\bibnamefont
  {Deng}}\ and\ \bibinfo {author} {\bibfnamefont {H.~W.~J.}\ \bibnamefont
  {Bl\"ote}},\ }\href {\doibase 10.1103/PhysRevE.70.035107} {\bibfield
  {journal} {\bibinfo  {journal} {Phys. Rev. E}\ }\textbf {\bibinfo {volume}
  {70}},\ \bibinfo {pages} {035107(R)} (\bibinfo {year} {2004})}\BibitemShut
  {NoStop}%
\bibitem [{\citenamefont {Qian}\ \emph {et~al.}(2005)\citenamefont {Qian},
  \citenamefont {Deng},\ and\ \citenamefont {Bl\"ote}}]{Qian2005}%
  \BibitemOpen
  \bibfield  {author} {\bibinfo {author} {\bibfnamefont {X.}~\bibnamefont
  {Qian}}, \bibinfo {author} {\bibfnamefont {Y.}~\bibnamefont {Deng}}, \ and\
  \bibinfo {author} {\bibfnamefont {H.~W.~J.}\ \bibnamefont {Bl\"ote}},\ }\href
  {\doibase 10.1103/PhysRevE.72.056132} {\bibfield  {journal} {\bibinfo
  {journal} {Phys. Rev. E}\ }\textbf {\bibinfo {volume} {72}},\ \bibinfo
  {pages} {056132} (\bibinfo {year} {2005})}\BibitemShut {NoStop}%
\bibitem [{\citenamefont {Deng}\ and\ \citenamefont
  {Bl\"ote}(2005)}]{Deng2005}%
  \BibitemOpen
  \bibfield  {author} {\bibinfo {author} {\bibfnamefont {Y.}~\bibnamefont
  {Deng}}\ and\ \bibinfo {author} {\bibfnamefont {H.~W.~J.}\ \bibnamefont
  {Bl\"ote}},\ }\href {\doibase 10.1103/PhysRevE.71.026109} {\bibfield
  {journal} {\bibinfo  {journal} {Phys. Rev. E}\ }\textbf {\bibinfo {volume}
  {71}},\ \bibinfo {pages} {026109} (\bibinfo {year} {2005})}\BibitemShut
  {NoStop}%
\bibitem [{\citenamefont {Deng}\ and\ \citenamefont
  {Bl\"ote}(2019)}]{Deng2019}%
  \BibitemOpen
  \bibfield  {author} {\bibinfo {author} {\bibfnamefont {Y.}~\bibnamefont
  {Deng}}\ and\ \bibinfo {author} {\bibfnamefont {H.~W.~J.}\ \bibnamefont
  {Bl\"ote}},\ }\href {\doibase 10.1103/PhysRevE.99.069904} {\bibfield
  {journal} {\bibinfo  {journal} {Phys. Rev. E}\ }\textbf {\bibinfo {volume}
  {99}},\ \bibinfo {pages} {069904(E)} (\bibinfo {year} {2019})}\BibitemShut
  {NoStop}%
\bibitem [{\citenamefont {Potts}(1952)}]{Potts1952}%
  \BibitemOpen
  \bibfield  {author} {\bibinfo {author} {\bibfnamefont {R.~B.}\ \bibnamefont
  {Potts}},\ }\href {\doibase 10.1017/S0305004100027419} {\bibfield  {journal}
  {\bibinfo  {journal} {Proceedings of the Cambridge Philosophical Society}\
  }\textbf {\bibinfo {volume} {48}},\ \bibinfo {pages} {106} (\bibinfo {year}
  {1952})}\BibitemShut {NoStop}%
\bibitem [{\citenamefont {Fateev}\ and\ \citenamefont
  {Zamolodchikov}(1987)}]{Fateev1987}%
  \BibitemOpen
  \bibfield  {author} {\bibinfo {author} {\bibfnamefont {V.}~\bibnamefont
  {Fateev}}\ and\ \bibinfo {author} {\bibfnamefont {A.}~\bibnamefont
  {Zamolodchikov}},\ }\href {\doibase
  https://doi.org/10.1016/0550-3213(87)90166-0} {\bibfield  {journal} {\bibinfo
   {journal} {Nuclear Physics B}\ }\textbf {\bibinfo {volume} {280}},\ \bibinfo
  {pages} {644 } (\bibinfo {year} {1987})}\BibitemShut {NoStop}%
\bibitem [{\citenamefont {Fuchs}\ and\ \citenamefont
  {Schweigert}(1998)}]{Fuchs1998}%
  \BibitemOpen
  \bibfield  {author} {\bibinfo {author} {\bibfnamefont {J.}~\bibnamefont
  {Fuchs}}\ and\ \bibinfo {author} {\bibfnamefont {C.}~\bibnamefont
  {Schweigert}},\ }\href {\doibase 10.1016/S0370-2693(98)01185-X} {\bibfield
  {journal} {\bibinfo  {journal} {Physics Letters B}\ }\textbf {\bibinfo
  {volume} {441}},\ \bibinfo {pages} {141 } (\bibinfo {year} {1998})},\ \Eprint
  {http://arxiv.org/abs/hep-th/9806121} {arXiv:hep-th/9806121} \BibitemShut
  {NoStop}%
\bibitem [{\citenamefont {Tang}\ \emph {et~al.}(2017)\citenamefont {Tang},
  \citenamefont {Chen}, \citenamefont {Li}, \citenamefont {Xie}, \citenamefont
  {Tu},\ and\ \citenamefont {Wang}}]{Tang2017}%
  \BibitemOpen
  \bibfield  {author} {\bibinfo {author} {\bibfnamefont {W.}~\bibnamefont
  {Tang}}, \bibinfo {author} {\bibfnamefont {L.}~\bibnamefont {Chen}}, \bibinfo
  {author} {\bibfnamefont {W.}~\bibnamefont {Li}}, \bibinfo {author}
  {\bibfnamefont {X.~C.}\ \bibnamefont {Xie}}, \bibinfo {author} {\bibfnamefont
  {H.-H.}\ \bibnamefont {Tu}}, \ and\ \bibinfo {author} {\bibfnamefont
  {L.}~\bibnamefont {Wang}},\ }\href {\doibase 10.1103/PhysRevB.96.115136}
  {\bibfield  {journal} {\bibinfo  {journal} {Phys. Rev. B}\ }\textbf {\bibinfo
  {volume} {96}},\ \bibinfo {pages} {115136} (\bibinfo {year} {2017})},\
  \Eprint {http://arxiv.org/abs/1708.04022} {arXiv:1708.04022
  [cond-mat.str-el]} \BibitemShut {NoStop}%
\bibitem [{\citenamefont {Zamolodchikov}\ and\ \citenamefont
  {Fateev}(1987)}]{Zamolodchikov1987}%
  \BibitemOpen
  \bibfield  {author} {\bibinfo {author} {\bibfnamefont {A.}~\bibnamefont
  {Zamolodchikov}}\ and\ \bibinfo {author} {\bibfnamefont {V.}~\bibnamefont
  {Fateev}},\ }\href {\doibase 10.1007/BF01028644h} {\bibfield  {journal}
  {\bibinfo  {journal} {Theoretical and Mathematical Physics}\ }\textbf
  {\bibinfo {volume} {71}},\ \bibinfo {pages} {451 } (\bibinfo {year}
  {1987})}\BibitemShut {NoStop}%
\bibitem [{\citenamefont {Evenbly}\ and\ \citenamefont
  {Vidal}(2016)}]{Evenbly2016}%
  \BibitemOpen
  \bibfield  {author} {\bibinfo {author} {\bibfnamefont {G.}~\bibnamefont
  {Evenbly}}\ and\ \bibinfo {author} {\bibfnamefont {G.}~\bibnamefont
  {Vidal}},\ }\href {\doibase 10.1103/PhysRevLett.116.040401} {\bibfield
  {journal} {\bibinfo  {journal} {Phys. Rev. Lett.}\ }\textbf {\bibinfo
  {volume} {116}},\ \bibinfo {pages} {040401} (\bibinfo {year} {2016})},\
  \Eprint {http://arxiv.org/abs/1510.00689} {arXiv:1510.00689
  [cond-mat.str-el]} \BibitemShut {NoStop}%
\end{thebibliography}%

\end{document}